\RequirePackage[hyphens]{url}
\documentclass[aps,twocolumn,showpacs,pra,amsmath,amssymb,floatfix,superscriptaddress]{revtex4-2}
\usepackage[colorlinks=true,citecolor=blue,linkcolor=RubineRed,urlcolor=blue]{hyperref} 
\usepackage[usenames,dvipsnames]{xcolor}
\usepackage{amsmath,amssymb,graphicx,color}
\usepackage{bm}
\usepackage{float}
\usepackage[caption=false]{subfig}
\usepackage[usenames,dvipsnames]{xcolor}
\usepackage[normalem]{ulem}
\usepackage{soul}
\usepackage[normalem]{ulem}
\usepackage{soul,xcolor}
\usepackage{mathrsfs}
\usepackage[hyphens]{url}

\usepackage[T1]{fontenc}

\usepackage{mathptmx} 

\begin{document}

\title{Exciting the long-lived Higgs mode in superfluid Fermi gases with particle removal}

	\author{Guitao Lyu}\email{guitao@zju.edu.cn}
    \affiliation{School of Physics and Zhejiang Institute of Modern Physics, Zhejiang University, Hangzhou, Zhejiang 310027, China}

	\author{Kui-Tian Xi} \email{xiphys@nuaa.edu.cn}
        \affiliation{College of Physics, Nanjing University of Aeronautics and Astronautics, Nanjing 211106, China}
	\affiliation{Key Laboratory of Aerospace Information Materials and Physics (NUAA), MIIT, Nanjing 211106, China}
        \affiliation{School of Physics and Zhejiang Institute of Modern Physics, Zhejiang University, Hangzhou, Zhejiang 310027, China}

	\author{Sukjin Yoon}
	\affiliation{Center for Theoretical Physics of Complex Systems, Institute for Basic Science, Daejeon 34051, Korea}
	\affiliation{Asia Pacific Center for Theoretical Physics, Pohang, Gyeongsangbuk-do 37637, Korea}
	\affiliation{Quantum Universe Center, Korea Institute for Advanced Study, Seoul 02455, Korea}

	\author{Qijin Chen}\email{qjc@ustc.edu.cn}
        \affiliation{Hefei National Research Center for Physical Sciences at the Microscale and School of Physical Sciences, University of Science and Technology of China,  Hefei, Anhui 230026, China}
        \affiliation{Shanghai Research Center for Quantum Science and CAS Center for Excellence in Quantum Information and Quantum Physics, University of Science and Technology of China, Shanghai 201315, China}
        \affiliation{Hefei National Laboratory, University of Science and Technology of China, Hefei 230088, China}

	\author{Gentaro Watanabe}\email{gentaro@zju.edu.cn}
	\affiliation{School of Physics and Zhejiang Institute of Modern Physics, Zhejiang University, Hangzhou, Zhejiang 310027, China}
	\affiliation{Zhejiang Province Key Laboratory of Quantum Technology and Device, Zhejiang University, Hangzhou, Zhejiang 310027, China}

	\date{\today}

\begin{abstract}
  Experimental evidence of the Higgs mode in strongly interacting superfluid Fermi gases had not been observed until recently [\href{https://doi.org/10.1038/s41567-018-0128-6}{Behrle \textit{et al.}, Nat. Phys. {\bf 14}, 781 (2018)}]. Due to the coupling with other collective modes and quasiparticle excitations, generating stable Higgs-mode oscillations is challenging. We study how to excite long-lived Higgs-mode oscillations in a homogeneous superfluid Fermi gas in the BCS-BEC crossover. We find that the Higgs mode can be excited by time-periodically modulating the scattering length at an appropriate amplitude and frequency. However, even for a modulation frequency below twice the pairing-gap energy, quasiparticles are still excited through the generation of higher harmonics due to nonlinearity in the superfluid. More importantly, we find that persistent Higgs-mode oscillations with almost constant amplitude can be produced by removing  particles at an appropriate momentum, and the oscillation amplitude  can be controlled by the number of removed particles. Finally, we propose two ways to experimentally realize of particle removal.
\end{abstract}

\maketitle

\section{Introduction}

The Higgs mode is an amplitude oscillation of the Higgs field which plays an important role in mass generation for elementary particles  via  spontaneous symmetry breaking \cite{Englert1964PRL, Higgs1964PRL, Guralnik1964PRL}. As an analogy, the Higgs mode in condensed-matter systems refers to an amplitude oscillation of the order parameter related to the spontaneous breaking of a continuous symmetry~\cite{Pekker2015AnnuRevCMP, Shimano2020AnnuRevCMP,Volovik2014JLTP}. While the excitation and detection of the Higgs mode in high-energy experiments require extremely large-scale facilities such as the Large Hadron Collider at CERN, they are feasible with tabletop experiments for condensed-matter systems~\cite{Sooryakumar1980PRL, Bissbort2011PRL, Endres2012Nature, Matsunaga2013PRL, Mansart2013PNAS, Sherman2015NatPhys, Hoang2016PNAS, Jain2017natphys, Hong2017NatPhys, Leonard2017Science, Behrle2018natphys}. Therefore, the latter is an important test bed for studying the Higgs mode and has been drawing attention especially after the detection of the Higgs boson in high-energy experiments in 2012~\cite{Aad2012PLB, Chatrchyan2012PLB}.

The first work on the Higgs mode in condensed matter was done by Volkov and Kogan around a half century ago \cite{Volkov1973JETP}, although it was not called the Higgs mode at that time. They found that the frequency of the Higgs mode in the superconductor is $\simeq 2\Delta_0/\hbar $ ($ \Delta_0 $ is the pairing gap in equilibrium) and the amplitude decays with time $t$ in a power law of $t^{-1/2}$. However, in the context of the BCS-BEC crossover in superfluid Fermi gases, it was predicted that the amplitude decay follows a different power law of $t^{-3/2}$ in the BEC regime~\cite{Gurarie2009PRL}, unlike  the $ t^{-1/2}$ behavior in the BCS regime. The earliest evidence for the existence of the Higgs mode was found in superconductors~\cite{Sooryakumar1980PRL}, where unexpected peaks in  Raman scattering were revealed to be the Higgs mode~\cite{Littlewood1981PRL, Littlewood1982PRB}. Since the realization of  degenerate atomic Fermi gases~\cite{Demarco1999science}, there have been many theoretical studies on the Higgs mode in superfluid Fermi gases with various types of external drive, such as a sudden quench~\cite{Andreev2004PRL, Barankov2004PRL, Szymanska2005PRL, Yuzbashyan2006PRL, Barankov2006PRL, Bulgac2009PRL, Gurarie2009PRL, Hannibal2015PRA, Yoon2017PRL, Hannibal2018PRA, Seibold2020PRB, Collado2022arXiv}, ramping~\cite{Yi2006PRA, Scott2012PRA, Tokimoto2019JPSJ}, time-periodic modulation of the interaction strength~\cite{Yi2006PRA, Scott2012PRA, Collado2018PRB, Collado2020PRB, Collado2021PRR}, and so on~\cite{Bruun2002PRL, Korolyuk2011PRA}. The visibility of the Higgs mode in fermionic superfluids has also been studied~\cite{Podolsky2011PRB, Liu2016PRA, Han2016PRA}. Regarding the small oscillations around the equilibrium state after  quench, a decay is inevitable in homogeneous three-dimensional systems in the BCS-BEC crossover since the Higgs mode is coupled with the Goldstone mode due to the absence of  particle-hole symmetry~\cite{Varma2002JLTP}.  On the other hand, when the initial perturbation is sufficiently large or for some particular nonequilibrium initial states, persistent Higgs-mode oscillations have been predicted to be possible~\cite{Barankov2004PRL, Yuzbashyan2006PRL, Barankov2006PRL, Bulgac2009PRL}.
Generating stable Higgs-mode oscillations is important for future practical applications. For example, persistent Higgs-mode oscillations enable us to probe the information about the material phase, such as the interband couplings in multiband superconductors (see, e.g., Refs.~\cite{Akbari2013EPL, Maiti2015PRB, Muller2018PRB}) and the superconductivity in photoinduced states (see, e.g., Refs.~\cite{Fausti2011science, Kaiser2014PRB, Hu2014naturemat, Mitrano2016nature}).
A long-lived Higgs mode is also possible in trapped two-dimensional Fermi gases, where the trapping confinement can stabilize the Higgs mode by making its decay channels discrete~\cite{Bruun2014PRA}. However,  experimental evidence for the Higgs mode in atomic superfluid Fermi gases was observed only in recent years \cite{Behrle2018natphys} because generating stable Higgs-mode oscillations against decay has been  challenging.

In the present paper, we propose a scheme for generating long-lived Higgs-mode oscillations in a homogeneous superfluid Fermi gas in the BCS-BEC crossover. We approach this problem by  numerical simulations of the time-dependent Bogoliubov-de Gennes (BdG) equations. First, we consider  time-periodic modulations of the scattering length in the BCS-BEC crossover. A basic idea behind this scheme is to sustain the Higgs-mode oscillations against decay by a continuous drive. Although this scheme can generate Higgs-mode oscillations, other unwanted excitations are also created due to the nonlinearity of the BdG equations. As an alternative scheme to excite the Higgs mode, we propose an unconventional type of quench by removing  particles around certain momentum. This scheme can generate stable and persistent Higgs-mode oscillations whose lifetime and amplitude are
much larger than those observed in a recent experiment~\cite{Behrle2018natphys}. By tuning the momentum of the removed particles at the peak of the pair wave function $F_k$, the long-lived Higgs mode can be excited. Furthermore, the amplitude of the oscillations can be controlled by the number of removed particles. The long lifetime and large amplitude are strong advantages for future experimental detections and investigations. Particularly, the amplitude of the oscillations excited by this scheme remains almost constant for a long time instead of experiencing prompt power-law decay.

This paper is organized as follows. In Sec.~\ref{sec:Formulation}, we present the formulation of our simulations. In Sec.~\ref{sec:harmonic excitations}, we discuss the scheme using the time-periodic modulation of the scattering length and its resulting higher-harmonic excitations. The scheme of removing  particles will be discussed in Sec.~\ref{sec:particle removal}. In Sec.~\ref{sec:conclusion}, we summarize our work and discuss how to experimentally realize the particle removal.

\section{Formulation}\label{sec:Formulation}

We consider a homogeneous superfluid Fermi gas at zero temperature in the BCS-BEC crossover. Our system consists of attractively interacting unpolarized (pseudo)spin-$1/2$ fermionic atoms. The time evolution of the pairing gap $\Delta (t)$ is studied by numerically solving the time-dependent Bogouliubov-de Gennes (TD-BdG) equations~\cite{deGennes1966book, Kummel1969ZPhysA, Ketterson1999book}:
\begin{equation}
  i \hbar \frac{d}{d t}
  \begin{bmatrix}
    u_{\bm{k}}(t) \\
    v_{\bm{k}}(t)
  \end{bmatrix}
  =
  \begin{bmatrix}
    H' & \Delta (t)  \\
    \Delta^{*} (t) & -H'
  \end{bmatrix}
  \begin{bmatrix}
    u_{\bm{k}}(t) \\
    v_{\bm{k}}(t)
  \end{bmatrix}\,,
  \label{eq:1}
\end{equation}
where $H' \equiv \hbar^2 k^{2} / (2m) - \mu = \epsilon_{k} - \mu$ is the single-particle Hamiltonian with wave number $k=|\bm{k}|$ and $\epsilon_{k} \equiv \hbar^2 k^2/2m$, $m$ is the mass of an atom, and $\mu$  is the chemical potential of the reference equilibrium state  introduced for convenience to remove the fast rotation of the overall phase. (In our simulations, the initial equilibrium state is taken as the reference state.) The time-dependent quasiparticle amplitudes $u_{\bm{k}}(t)$ and $v_{\bm{k}}(t)$ satisfy the normalization condition: $|u_{\bm{k}}(t)|^2 +|v_{\bm{k}}(t)|^2 = 1$.
The pairing gap reads
\begin{equation}
  \Delta(t) = -\frac{g}{V} \sum_{\bm{k}} u_{\bm{k}}(t)\, v_{\bm{k}}^{*}(t) = -\frac{g}{V} \sum_{\bm{k}} F_{\bm{k}}(t) \,,\label{eq:pairing gap}
\end{equation}
where $g$ is the coupling constant of the contact interaction and $V$ is the volume of the system. Here,
\begin{equation}
  F_{\bm{k}}(t) \equiv u_{\bm{k}}(t)\, v_{\bm{k}}^*(t)\,
\end{equation}
is the pair wave function. The number density $n$ of atoms is
\begin{align}
  n = \frac{2}{V} \sum_{\bm{k}} \left|v_{\bm{k}} (t)\right|^2 = \frac{2}{V} \sum_{\bm{k}} n_{\bm{k}}(t)\,,\label{eq:number}
\end{align}
with
\begin{align}
  n_{\bm{k}}(t) \equiv \left|v_{\bm{k}} (t)\right|^2
\end{align}
being the momentum distribution of atoms (either spin up or spin down). The energy density $\mathcal{E}$ is given by
\begin{equation}
  \mathcal{E} = \frac{2}{V} \sum_{\bm{k}} \epsilon_{k} \left|v_{\bm{k}}(t)\right|^2 + \frac{1}{g} \left|\Delta (t)\right|^2\,.
\end{equation}
All the summations here are restricted to the energy range $ 0 \le \epsilon_{k} \le E_{c}$, where $E_c \equiv \hbar^2k_c^2/2m$ is the cutoff energy ($k_c$ is the cutoff wave number), and the coupling constant $g$ is renormalized as~\cite{Leggett1980proc, Randeria1995book, Giorgini2008RMP}
\begin{align}
  \frac{1}{g} = \frac{m}{4\pi\hbar^2 a} - \frac{1}{V}\sum_{k \le k_c} \frac{1}{2\epsilon_{k}} = \frac{m}{4\pi\hbar^2 a} - \frac{mk_c}{2 \pi^2 \hbar^2}\,,\label{eq:reg}
\end{align}
where $a$ is the $s$-wave scattering length.

In our numerics,  we first solve self-consistently the gap equation (\ref{eq:pairing gap}) and the number equation (\ref{eq:number}) for the equilibrium solution for $\{u_{\bm k}\}$ and $\{v_{\bm k}\}$, by tuning the chemical potential in an iterative way. Then we treat the equilibrium solution as the initial state, and solve the TD-BdG equations (\ref{eq:1}) in momentum space using the fifth-order Adams–Bashforth backward predictor-corrector method. In each step of the time integration, the corrector step is iterated until the absolute error of all $u_{\bm k}$ and $v_{\bm k}$ is less than $10^{-6}$. The step size $\delta k$ of the momentum grid is taken to be $\delta k=0.001k_F$. The time step $\delta t$ is typically taken to be  $\delta t=0.0005 \hbar/E_F $, but is allowed to be adjusted to reach convergence.

\section{Higher harmonic excitations from time-periodic modulation}\label{sec:harmonic excitations}

First, we consider time-periodic modulation of the coupling constant $g$, which can be experimentally achieved by modulating the magnetic field near a Feshbach resonance~\cite{Greiner2005PRL, Thompson2005PRL, Clark2017nature} (see also Ref.~\cite{Chin2010RMP} and references therein). The coupling constant $g$ is periodically modulated around the initial value $g_0$ as
\begin{equation}
  g(t) = g_{0} \left( 1 + A \sin \omega t \right),
  \label{eq:modulation}
\end{equation}
where $A$ and $\omega$ are the modulation amplitude and frequency, respectively.

In the simulation, we initially prepare the equilibrium state at zero temperature for a given value of the interaction strength $1/k_Fa$, where $k_F=(3\pi^2 n)^{1/3}$ is the Fermi wave number of a uniform ideal Fermi gas with the same density $n$. From Eq.~(\ref{eq:reg}), this given value of $1/k_Fa$ and the corresponding initial coupling constant $g_0$ are related as
\begin{equation}
  \dfrac{1}{k_Fa}=\frac{4\pi\hbar^2}{m}\frac{1}{k_Fg_0} + \frac{2}{\pi} \sqrt{\frac{E_c}{E_F}}\,
\end{equation}
with $E_F \equiv \hbar^2 k_F^2/2m$. Starting from this initial equilibrium state, we solve the TD-BdG equations (\ref{eq:1}) under the time-periodic modulation of the coupling constant given by Eq.~(\ref{eq:modulation}).

By diagonalizing the matrix on the right-hand side of Eq.~(\ref{eq:1}) for the stationary state, one can readily obtain the quasiparticle spectrum $\pm E_k$ in the equilibrium state given by
\begin{align}
  E_k = \sqrt{(\epsilon_{k} - \mu)^2 + |\Delta_0|^2}\,.\label{eq:ek}
\end{align}
Here, $\Delta_0$ is the equilibrium value of the pairing gap, which can be determined self-consistently from Eqs.~(\ref{eq:pairing gap}) and (\ref{eq:number}).

\subsection{Threshold frequency in the small-amplitude limit}

\begin{figure}[t]
  \centering  \includegraphics[width=1\linewidth]{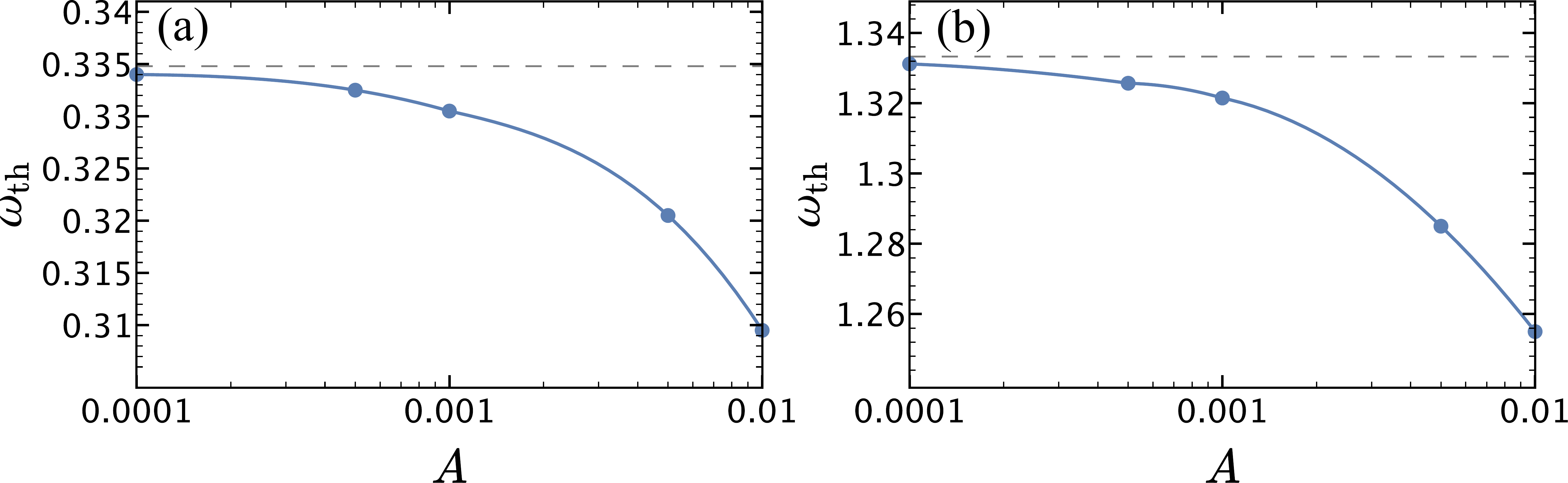}
  \caption{Threshold frequency $\omega_{\rm th}$ (in units of $E_F/\hbar$) as a function of the modulation amplitude $A$ for (a) $1/k_Fa=-1$ and (b) $0$. The horizontal dashed lines show $2|\Delta_0|/\hbar$, where $\Delta_0$ is the pairing gap in the equilibrium state for the same value of $1/k_Fa$.
  }
  \label{fig:threshold}
\end{figure}

Let us first focus on the small modulation amplitude $A \ll 1$. When the modulation frequency $\omega$ is sufficiently small such that $\hbar\omega \ll |\Delta|$, the system follows the modulation adiabatically, and the energy density $\mathcal{E}$ just shows a small amplitude oscillation due to the time-periodic change in the coupling constant $g(t)$. In this case, there are no excitations created. However, above a threshold frequency $\omega_{\rm th}$, i.e., $\omega \ge \omega_{\rm th}$, the energy density $\mathcal{E}$ shows a gradual increase in time in addition to the small amplitude oscillation. Figure \ref{fig:threshold} shows $\omega_{\rm th}$ as a function of $A$ for $1/k_Fa=-1$ [Fig.~\ref{fig:threshold} (a)] and $0$ [Fig.~\ref{fig:threshold} (b)]. Here, we see that, for nonzero modulation amplitude $A$, $\omega_{\rm th}$ is slightly smaller than $2 |\Delta_0|/\hbar$ (the dashed lines) corresponding to the threshold energy of the pair-breaking excitation for the equilibrium system (note that $\mu>0$ for both $1/k_Fa=-1$ and $0$). This is due to the reduction in the minimum value of $|\Delta(t)|$ from $\Delta_0$ directly caused by the reduction in the coupling constant $g(t)$ from $g_0$ to $g_0 (1-A)$ for nonzero $A$. Note that $\omega_{\rm th}$ approaches $2 |\Delta_0|/\hbar$ in the limit of $A\rightarrow 0$, as expected. On the other hand, in the sufficiently deep BEC regime with $\mu<0$ (e.g., at $1/k_Fa = 1$), the threshold frequency $\omega_{\rm th}$ is $2\sqrt{\mu^2+|\Delta_0|^2}/\hbar$ in the limit of $A\rightarrow 0$, as can be seen from Eq.~(\ref{eq:ek}).

\subsection{Resonance peaks and higher-harmonic generation\label{sec:highharm}}

\begin{figure}[t]
 \centering  \includegraphics[width=1\linewidth]{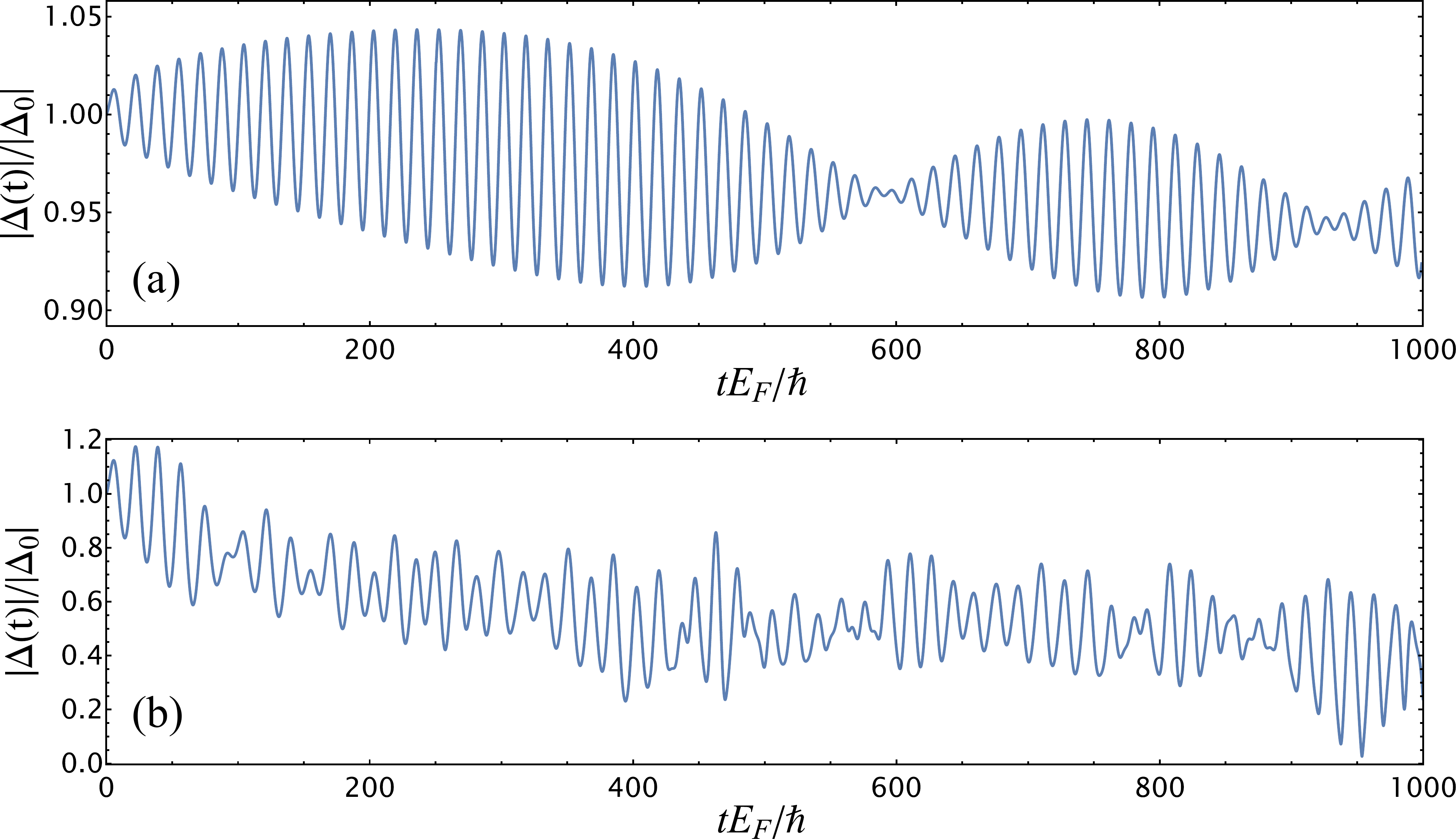}
  \caption{Oscillation of the pairing gap $|\Delta(t)|$ driven by the time-periodic modulation of the coupling constant $g(t)$ given by Eq.~(\ref{eq:modulation}). Here we take the modulation frequency $\omega=2|\Delta_0|/\hbar$ and $1/k_Fa = -1$. In (a) and (b), we take the modulation amplitude $A=0.001$ and $0.01$, respectively.
  }
  \label{fig:beating}
\end{figure}

Now, we employ the time-periodic modulation of $g$ to excite the Higgs mode. By tuning the modulation frequency $\omega$ around $2|\Delta_0|/\hbar$ just above $\omega_{\rm th}$ \cite{Volkov1973JETP}, we can resonantly excite the oscillation of the magnitude of the pairing gap $\Delta(t)$. Figure \ref{fig:beating}(a) shows an example of the resulting pairing dynamics induced by the time-periodic modulation of $g$ with $\omega=2|\Delta_0|/\hbar$ and very small amplitude $A=0.001$. Here, we can see that, in addition to the oscillation with a constant frequency at $\simeq 2|\Delta_0|/\hbar$, which corresponds to the Higgs mode, there is a clear signature of  beating, and the amplitude of its envelope decreases in time. Such a diminishing beat is evidence that the time-periodic modulation of $g$ excites not only the Higgs mode but also other unwanted collective modes and quasiparticles.

Furthermore, in addition to the collective modes, various single-particle excitations are also created by the modulation of $g$ at a larger amplitude $A$ and/or after a sufficiently long time. One prominent phenomenon is the higher-harmonic generation by the parametric up-conversion discussed below (see also Ref.~\cite{Collado2020PRB}). Due to these excitations, the oscillation of $|\Delta(t)|$ becomes irregular, as shown in Fig.~\ref{fig:beating}(b) for $A=0.01$. To see the mechanism of  higher-harmonic generation, we first write the pairing gap $\Delta(t)$ as $\Delta(t) = g(t)\, \tilde{\Delta}(t)$, with $ \tilde{\Delta}(t) \equiv -V^{-1} \sum_{\bm k} u_{\bm k}(t)\, v_{\bm k}^*(t)$. Suppose that $\tilde{\Delta}(t)$ initially synchronizes with the modulation of the coupling constant $g(t)$, so that it oscillates at the modulation frequency $\omega$. Since $g(t)$ also oscillates at $\omega$, their product $g(t)\, \tilde{\Delta}(t)$ has the second-harmonic component with doubled frequency $2\omega$. Once $\tilde{\Delta}(t)$ has the component of $2\omega$, $g(t)\, \tilde{\Delta}(t)$ will have the third-harmonic component with $3\omega$. Likewise, further higher-harmonic excitations are also generated. This phenomenon is analogous to the parametric up-conversion in nonlinear optics (see, e.g., \cite{Yariv1988book}).

\begin{figure}[t]
  \centering  \includegraphics[width=0.6\linewidth]{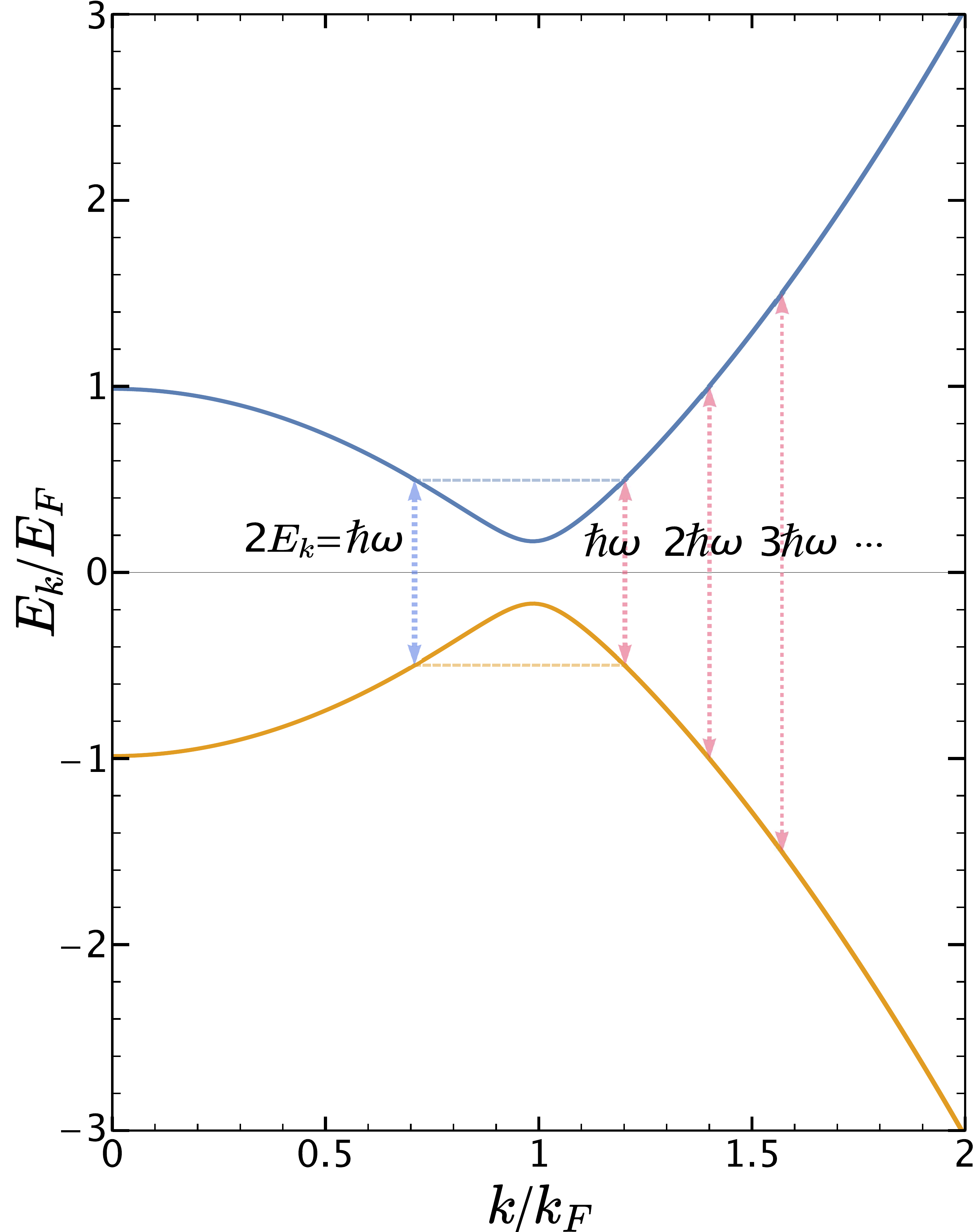}
  \caption{Quasiparticle spectra $\pm E_k$ and positions of the resonance peaks for the fundamental and higher harmonics. This plot is shown for $1/k_Fa=-1$ and $\omega=1 E_F/\hbar$ as an example.
  }
  \label{fig:spectra}
\end{figure}

\begin{figure}[t]
  \centering  \includegraphics[width=1\linewidth]{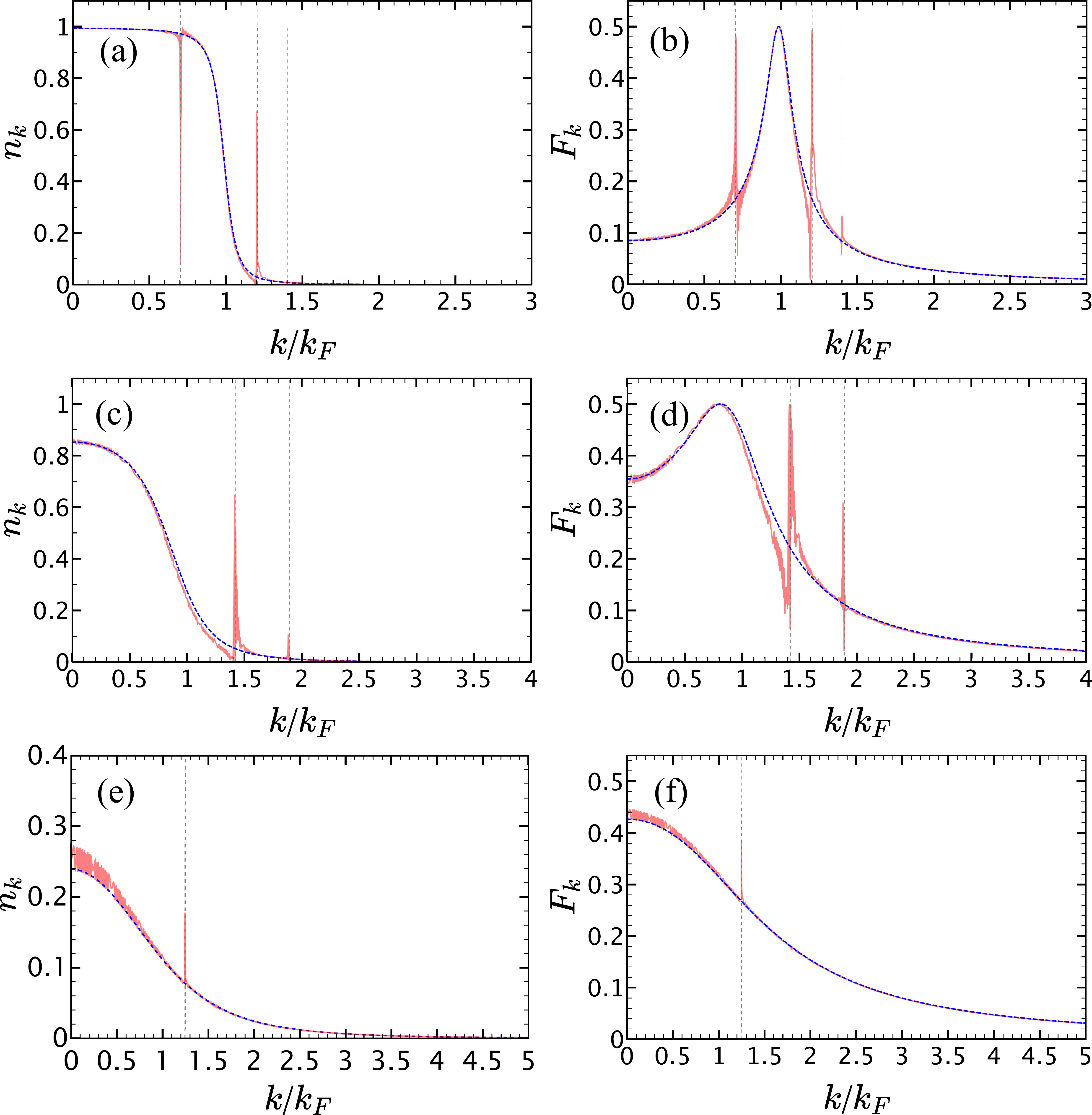}
  \caption{Resonance peaks in $n_k$ and $F_k$ for $1/k_Fa=-1$ [(a) and (b)], $1/k_Fa=0$ [(c) and (d)], and $1/k_Fa=1$ [(e) and (f)]. The red solid line shows the snapshot at $t=2000/(E_F/\hbar)$ and the blue dashed line shows the initial equilibrium state. Here, the modulation amplitude $A=0.01$. For (a) and (b), the modulation frequency is $\omega=1E_F/\hbar$, and the vertical dotted lines are obtained from the conditions $2E_k = \hbar\omega$, $\hbar\omega$, and $2\hbar\omega$, respectively (from left to right). For (c) and (d), the modulation frequency is $\omega=3E_F/\hbar$, and the vertical dotted line is obtained from the conditions $2E_k = \hbar\omega$ and $2\hbar\omega$, respectively (from left to right). For (e) and (f), the modulation frequency is $\omega=3E_F/\hbar$, and the vertical dotted line is obtained from the condition $2E_k = 2\hbar\omega$. }
  \label{fig:res_kfainv}
\end{figure}

Due to the higher-harmonic generation, we can observe growing peaks (and dips) in the momentum distribution $n_k$ and the pair wave function $F_k$ during the time evolution. Furthermore, even if the modulation frequency is smaller than the threshold frequency $\omega_{\rm th} \alt 2|\Delta_0|/\hbar$ for $\mu \ge 0$ or $\omega_{\rm th} \alt 2\sqrt{\mu^2 + |\Delta_0|^2}/\hbar$ for $\mu < 0$, resonance peaks and dips can appear by the higher harmonics. For example, Fig.~\ref{fig:res_kfainv} shows the snapshots of $n_k$ and $F_k$ at $t=2000/(E_F/\hbar)$ from the BCS (top panels) to BEC (bottom panels) regime. The positions of these peaks (and dips) are consistent with the resonance condition (see Fig.~\ref{fig:spectra}):
\begin{equation}
  2E_k = l \hbar\omega\quad (l=1, 2, 3, \cdots)\,,
  \label{eq:res}
\end{equation}
with the quasiparticle spectrum $E_k$ given by Eq.~(\ref{eq:ek}).

Note that, when $\mu>0$ so that $E_k$ is not monotonically increasing with $k$, modulation with $\omega$ satisfying
\begin{align}
  2|\Delta_0|< l\hbar\omega < 2\sqrt{\mu^2+|\Delta_0|^2}\label{eq:multires}
\end{align}
yields two resonance peaks (and dips) with the same value of $l$. See, e.g., the two vertical arrows for $2E_k=\hbar \omega$ in Fig.~\ref{fig:spectra}. As an example, Figs.~\ref{fig:res_kfainv}(a) and \ref{fig:res_kfainv}(b) show the case of $1/k_Fa=-1$ (BCS side, $\mu>0$) with $\omega=1 E_F/\hbar$, which satisfies Eq.~(\ref{eq:multires}) for $l=1$ ($\mu \approx 0.95 E_F$ and $\Delta_0 \approx 0.21 E_F$ at $1/k_Fa=-1$ in the mean-field theory). There are two peaks/dips in $n_k$ and $F_k$ at $k/k_F\approx 0.7$ and $1.2$, and the positions of these peaks/dips are consistent with the resonance condition (\ref{eq:res}) (marked by the vertical dotted line) for $l=1$. There is another peak/dip at $k/k_F\approx 1.4$, which corresponds to the resonance condition (\ref{eq:res}) for $l=2$. Figures~\ref{fig:res_kfainv}(c) and \ref{fig:res_kfainv}(d) show the case at unitarity ($1/k_Fa=0$) with $\omega= 3E_F/\hbar$. 
\begin{figure}[t]
  \centering  \includegraphics[width=1\linewidth]{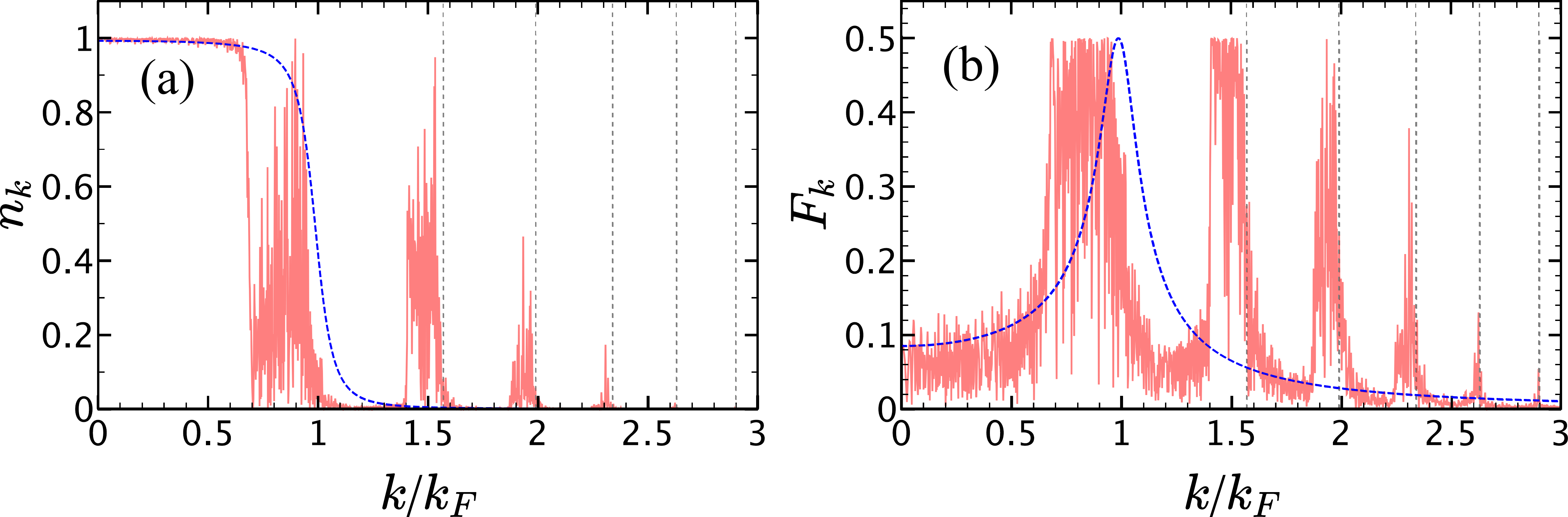}
  \caption{Resonance peaks in (a) $n_k$ and (b) $F_k$ for large-amplitude modulation with $A=0.3$, for $1/k_Fa=-1$ and $\omega=3E_F/\hbar$. The red solid line shows the snapshot at $t=2000/(E_F/\hbar)$, and the blue dashed line shows the initial equilibrium state. The vertical dotted lines correspond to $2E_k = \hbar\omega,\ 2\hbar\omega,\ 3\hbar\omega,\ 4\hbar\omega$, and $5\hbar\omega$, respectively (from left to right), in both panels.
  }
  \label{fig:res_largeamp}
\end{figure}
Although $\mu>0$ at unitarity, this value of $\omega$ does not satisfy Eq.~(\ref{eq:multires}) for any integer $l$ ($\mu \approx 0.59 E_F$ and $\Delta_0 \approx 0.69 E_F$ at $1/k_Fa=0$ in the mean-field theory), so that there is only one peak/dip for each value of $l$: there are two peaks in $n_k$ and $F_k$ at $k/k_F\approx 1.4$ and $1.9$, which correspond to the fundamental harmonic ($l=1$) and the second harmonic ($l=2$), respectively. Figures~\ref{fig:res_kfainv}(e) and \ref{fig:res_kfainv}(f) show the case for $1/k_Fa=1$ as an example on the BEC side. Since $\mu<0$ at $1/k_Fa=1$, the quasiparticle spectrum $E_k$ is monotonically increasing with $k$, so that the resonance condition (\ref{eq:res}) for each value of $l$ can yield only a single peak and dip. As an representative example, we set $\omega = 3E_F/\hbar$ smaller than the threshold frequency $\omega_{\rm th} \alt 2\sqrt{\mu^2 + \Delta^2} \approx 3.11 E_F/\hbar$ for $\mu \approx -0.80 E_F$ and $\Delta_0 \approx 1.33 E_F$ at $1/k_Fa=1$. Note that, even though $\omega < \omega_{\rm th}$, there is a resonance peak around $k/k_F \approx1.24$ which is given by the second harmonic $l=2$.

For larger modulation amplitudes $A$, resonance peaks and dips grow much faster. Figure~\ref{fig:res_largeamp} shows the result for $A=0.3$ at $t=2000/(E_F/\hbar)$, where many resonance peaks in $n_k$ and $F_k$ caused by the higher harmonics can clearly be seen. In this example, we set $\omega=3E_F/\hbar$, so that the first resonance peak of $l=1$ is located far above the Fermi wave number, $k\approx 1.57k_F$. Since many particles are excited to the resonance peaks above the Fermi energy, the distribution $n_k$ of the particles below the Fermi energy is largely depleted from the initial equilibrium state shown by the blue dashed line.

\section{Long-lived Higgs mode excited by particle removal} \label{sec:particle removal}
In this section, we propose an alternative scheme to excite long-lived Higgs-mode oscillations. This scheme consists of removing particles around  a certain magnitude of the momentum (denoted by $\hbar k_{\text{hole}}$), which can be realized by, e.g., shining red-detuned lasers, which will be discussed in detail in  the next section. Removing  particles can be regarded as an alternative to the conventional way of quenching though parameter change. As we shall demonstrate below, an advantage of this scheme is that very long-lived Higgs-mode oscillations with relatively large amplitude can be produced in a wide region in the crossover from the BCS limit ($1/k_Fa \ll -1$) to the unitary regime ($1/k_F \approx 0$).

\begin{figure}[tb]
	\centering  \includegraphics[width=1\linewidth]{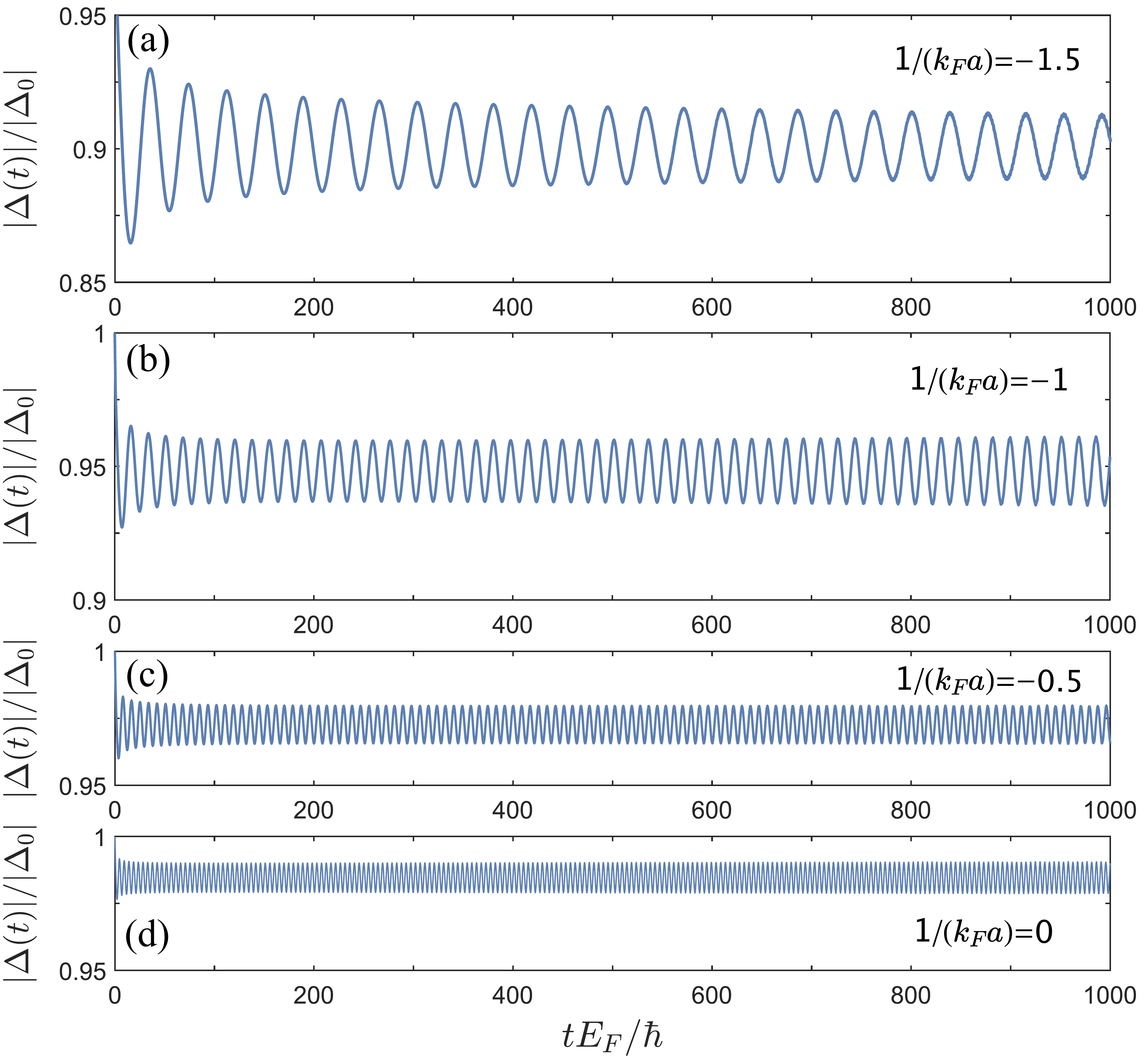}
	\caption{Long-lived Higgs-mode oscillations excited by removing  particles in superfluid Fermi gases with different $1/k_F a$. $5\%$ of the particles are removed around the wave number $k_{\text{hole}}$ at $k_{\text{hole}}/k_F=0.989$, $0.975$, $0.93$, and $0.79$ in (a)--(d), respectively. Here, we take $E_c =100 E_F$.
	}
	\label{fig:dn0_05_oscillation}
\end{figure}

Figure~\ref{fig:dn0_05_oscillation} shows the resulting long-lived Higgs-mode oscillations excited using this scheme at several values of $1/k_Fa$. In the simulation, we first prepare the equilibrium state for a given value of $1/k_Fa$, and then remove the particles around the wave number $k_{\text{hole}}$. For the cases of Figs.~\ref{fig:dn0_05_oscillation}(a)--\ref{fig:dn0_05_oscillation}(d), we set $k_{\text{hole}}/k_F=0.989, \ 0.975,\ 0.93$, and $0.79$, respectively, and the number of removed particles $\delta n  V$ is $5\%$ of the total number of particles (i.e., $\delta n/n = 5\%$). Specifically, in the simulations, we remove the particles in a finite range of $k$ around $k_{\text{hole}}$ with the width $2\delta k$, i.e., $k_{\text{hole}}-\delta k \le k \le k_{\text{hole}}+\delta k$. The width $2\delta k$ is determined by the number of removed particles based on the following condition:
\begin{align}
  \delta n = {2} \int_{|k-k_{\text{hole}}| \le \delta k}  n_k\, d^3 k
  \label{eq:delta_n}
\end{align}
with $n_k$ for the initial equilibrium state. The oscillation frequency of $|\Delta(t)|$ in each panel is around $2\Delta_0/\hbar$, which agrees with the Higgs mode frequency. It is noted that, after the oscillation amplitude stabilizes, $|\Delta(t)|$ does not show a decaying oscillation [see, e.g., Figs.~\ref{fig:dn0_05_oscillation}(b)--\ref{fig:dn0_05_oscillation}(d)]. Rather, the amplitude even gradually increases after the stabilization: in a very long timescale, the amplitude of the oscillation gradually increases to a maximum value and then decreases to a nonzero minimum value repeatedly. The time $t_1$ corresponding to the first maximum amplitude, which characterizes the duration of the constant amplitude, serves as a reasonable measure for the effective lifetime of the Higgs mode in the present case.

\begin{figure*}[tbh]
  \centering  \includegraphics[width=0.85\linewidth]{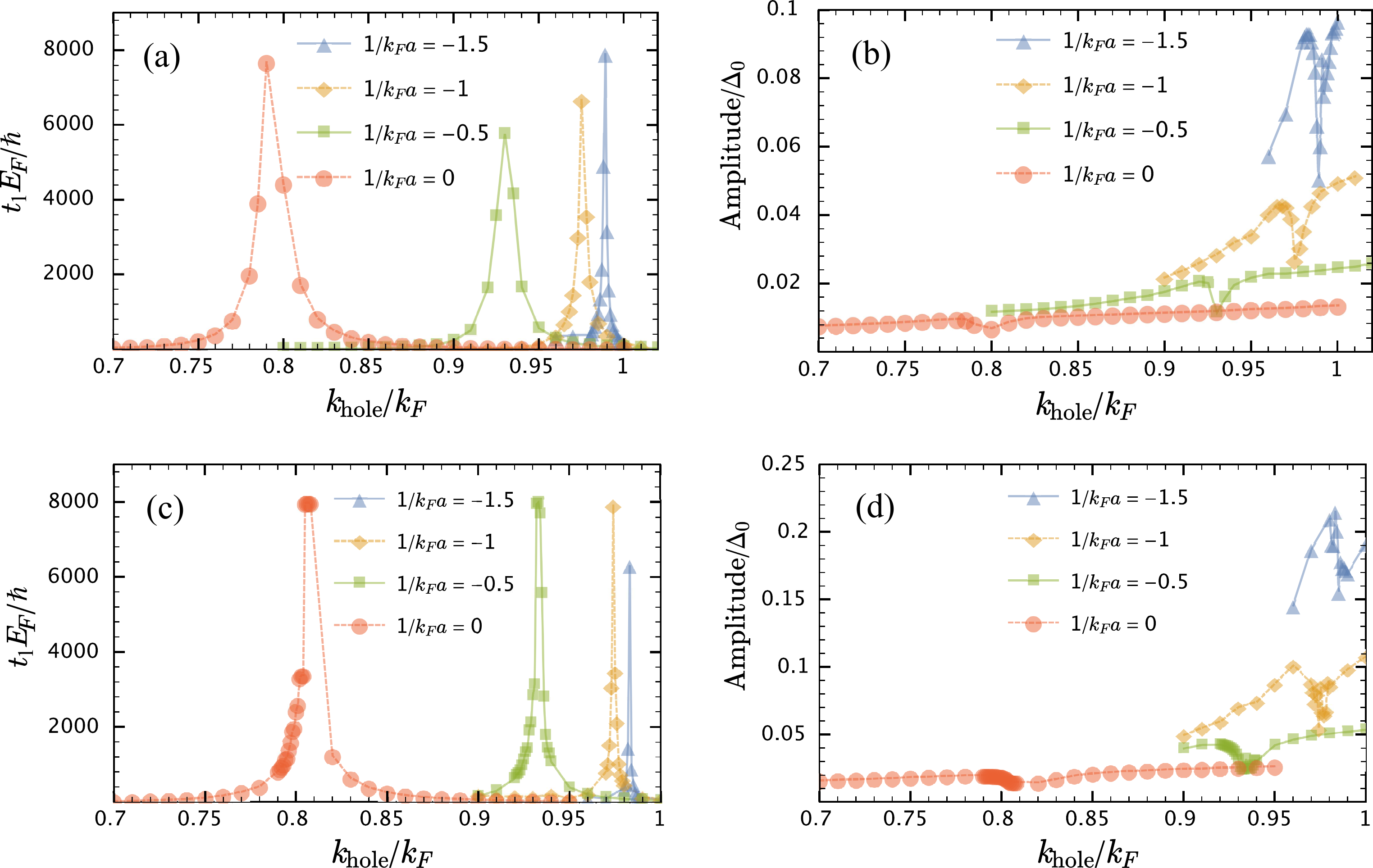}
  \caption{Time $t_1$ at which the oscillation amplitude takes the maximum value for the first time [(a) and (c)] and the relative amplitude at $t_1$ [(b) and (d)] as a function of $k_{\text{hole}}$ for different values of $1/k_Fa$. The number of removed particles is $\delta n/n = 5\%$ in (a) and (b) and $10\%$ in (c) and (d). The cutoff energy $E_c$ is taken at $100 E_F$ for all the cases in (a) and (b). In (c) and (d), we take $E_c = 16E_F$ for $1/k_Fa = -1.5$ and $-1$, and $E_c = 36E_F$ for $1/k_Fa = -0.5$ and $0$.
  }
  \label{fig:k_hole}
\end{figure*}

We study the Higgs mode excitation by removing  particles at various values of $k_{\text{hole}}$ for different $1/k_Fa$. The results are summarized in Fig.~\ref{fig:k_hole}: Figs.~\ref{fig:k_hole}(a) and \ref{fig:k_hole}(b) show the case of $\delta n/n = 5\%$, and Figs.~\ref{fig:k_hole}(c) and \ref{fig:k_hole}(d) show $\delta n/n = 10\%$. Figure~\ref{fig:k_hole}(a) shows the time $t_1$ as a function of $k_{\text{hole}}$. We observe that the long-lived Higgs mode is excited only in a certain range of $k_{\text{hole}}$, and the range of $k_{\text{hole}}$ changes with $1/k_F a$. In the unitary regime ($1/k_F a \approx 0$), the range of $k_{\text{hole}}$ to excite the long-lived Higgs mode is wider than that in the BCS regime ($1/k_F a < 0$). The long-lived Higgs-mode oscillations shown in Fig.~\ref{fig:dn0_05_oscillation} are realized for a particular value of $k_{\text{hole}}$ corresponding to the resonance peak in Fig.~\ref{fig:k_hole}(a).

Figure~\ref{fig:k_hole}(b) shows the maximum amplitude at $t_1$ relative to $\Delta_0$ as a function of $k_{\text{hole}}$. The maximum relative amplitude becomes larger by going towards the deeper BCS regime. On the other hand, it is hard to excite a Higgs-mode oscillation with a larger amplitude in the BEC regime ($1/k_Fa > 0$, not shown) using the present scheme. Note that the relative amplitude shows a dip at the wave number $k_{\text{hole}}$, corresponding to the resonance peak location in Fig.~\ref{fig:k_hole}(a). This dip can be understood as follows. When we have a long-lived Higgs-mode oscillation with  $t_1 = \infty$ at $k_{\text{hole}}$  corresponding to the resonance peak location, no other modes except for the Higgs mode are excited. However, if $k_{\text{hole}}$ slightly deviates from the resonance, not only the Higgs mode but other modes are also excited, so that the oscillation resulting from the superposition of these different excited modes has a larger amplitude compared to the former case of only the Higgs-mode excitation.

Next, we compare the cases with different fractions of removed particles. Figures~\ref{fig:k_hole}(c) and \ref{fig:k_hole}(d) are the same as Figs.~\ref{fig:k_hole}(a) and \ref{fig:k_hole}(b), respectively, except for $\delta n/n = 10\%$. In comparison of Figs.~\ref{fig:k_hole}(a) and \ref{fig:k_hole}(b) for $\delta n/n = 5\%$ and Figs.~\ref{fig:k_hole}(c) and \ref{fig:k_hole}(d) for $\delta n/n = 10\%$, the effective lifetime $t_1$ is almost the same in the two cases, but the amplitude of the oscillation for $\delta n/n = 10\%$ is roughly double that for $\delta n/n = 5\%$ [see Figs.~\ref{fig:k_hole}(b) and \ref{fig:k_hole}(d)]. This means that the amplitude of the excited Higgs-mode oscillation can be controlled by tuning the number of removed particles $\delta n/n$ in the present scheme. It should also be noted that, even in the case of $\delta n/n=5\%$, the relative amplitude, which is of the order of $1 \%$ [see Fig.~\ref{fig:k_hole}(b)], is much larger than that in the recent experiment \cite{Behrle2018natphys}, which was only of the order of $0.1\%$.

Finally, we remark that the wave number $k_{\text{hole}}$ of the resonance peak in Figs.~\ref{fig:k_hole}(a) and \ref{fig:k_hole}(c) coincides with the wave number $k_{\rm peak}$ at the peak location of the pair wave function $F_k$ in the equilibrium state [see, e.g., the blue dashed lines in Figs.~\ref{fig:res_kfainv}(b) and \ref{fig:res_kfainv}(d)]. Thus, one can deduce from  the pair wave function $F_k$ the appropriate value of $k_{\text{hole}}$ for exciting the Higgs mode. Since the peak of $F_k$ at nonzero $k$ disappears in the BEC regime of $1/k_Fa \agt 0.5$, our scheme is applicable for exciting the long-lived Higgs-mode oscillation up to the shallow BEC side of $1/k_F a \alt 0.5$.

\section{Summary and concluding remarks} \label{sec:conclusion}

Finally, we shall provide two possible methods for  experimental realization of  particle removal. The first method is to use resonant laser beams to transfer atoms with certain momenta to an untrapped internal state. Taking a quasi-one-dimensional Fermi gas, e.g., an elongated Fermi gas along the $x$ direction in a tight trap in the $y$ and $z$ directions, we shine two counterpropagating red-detuned lasers with wave number $q$ and detuning $\delta$ in the $\pm x$ directions to resonantly excite the atoms around certain momenta $\pm \hbar k_x = \pm m \delta/q$. Here, the excited state is supposed to be an untrapped state of the external potential. Note that the untrapped excited state should have a narrow natural linewidth to make the momentum selection accurate enough compared to the width $\Delta k_{\rm hole}$ of the peaks in Figs.~\ref{fig:k_hole}(a) and \ref{fig:k_hole}(c) [typically, $\Delta k_{\rm hole} = O(10^{-2})\,k_F$]. The clock states [usually with a natural linewidth on the order of $O(10)$~mHz] can fulfill this requirement very well. For example, there is a clock state with a linewidth of $38.5$~mHz for $ ^{173}$Yb \cite{Porsev2004PRA}. A degenerate Fermi gas of $^{173}$Yb realized in Ref.~\cite{Fukuhara2007PRL} has a Fermi temperature $T_F \approx 2 \times 10^2$~nK, corresponding to the Fermi energy $E_F / h \approx 4$~kHz. Therefore, the linewidth of the clock state is negligible compared to the energy width $\Delta E_{\rm hole}$ corresponding to the region within $\Delta k_{\rm hole}$ around $k_{\rm hole}$: $\Delta E_{\rm hole} \sim \hbar^2 (k_{\rm hole}+ \Delta k_{\rm hole})^2/2m - \hbar^2 k_{\rm hole}^2/2m \simeq 2 \hbar^2 k_{\rm hole}\, \Delta k_{\rm hole}/2m \sim O(10^2)$~Hz$\times h$ for $k_{\rm hole} \sim k_F$ and $\Delta k_{\rm hole} = O(10^{-2})\,k_F$. Thus, the momentum can be selected accurately provided the laser frequency is well locked.

\begin{figure}[tb]
  \centering  \includegraphics[width=1\linewidth]{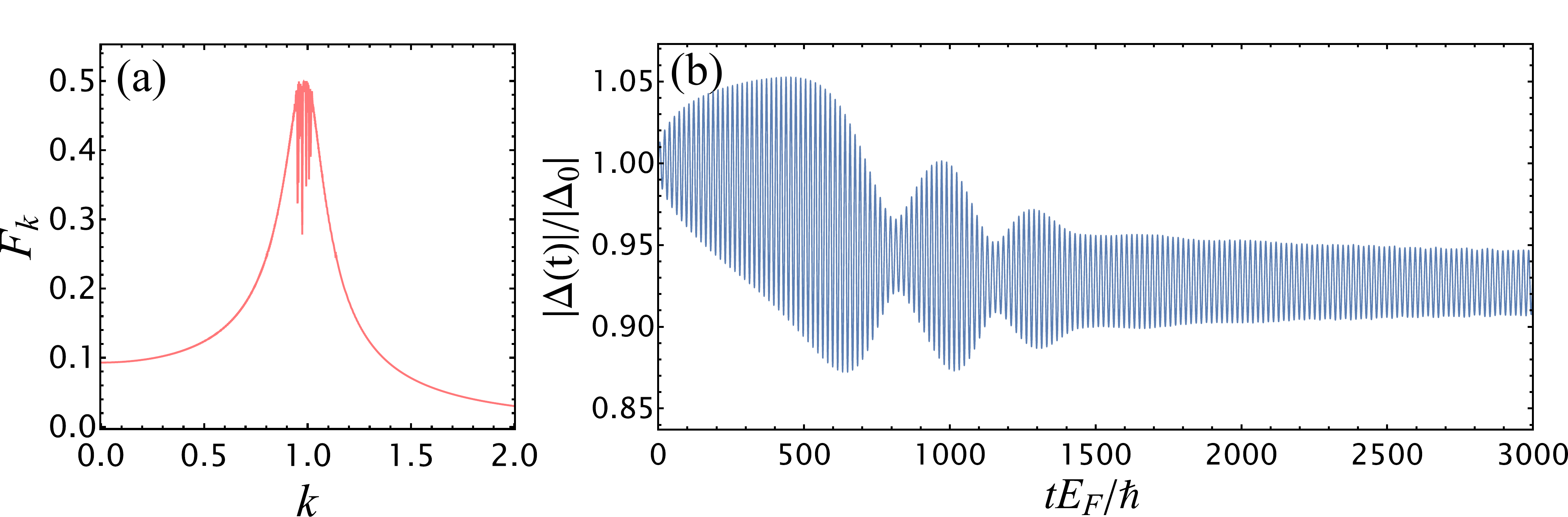}
  \caption{A Higgs-mode oscillation by removing  particles using $g$ modulation. Here, we take $1/k_F a=-1$ and the modulation frequency $\omega=1.97\,|\Delta_0|/\hbar$. The modulation amplitude is constant at $A=0.001$ from $ t = 0$ to $1000\,\hbar/E_F$, and then reduced to $0$ linearly from $ t = 1000\,\hbar/E_F$ to $1500\,\hbar/E_F$. Panel (a) shows $F_k$ at $ t = 1500\,\hbar/E_F$ with a dip around $k_{\rm peak}$. Panel (b) shows the oscillation of the magnitude of the order parameter $\Delta (t)$ during the whole process. Here we take $E_c =100\,E_F$.
  }
  \label{fig:g_modulation_at_peak}
\end{figure}

The second method is to make the $g$ modulation (see Sec.~\ref{sec:highharm}) with frequency $\omega \simeq 2 \Delta_0/\hbar$ to make a resonance at the wave number $k_{\rm peak}$ corresponding to the peak of the pair wave function $F_k$. Here, we keep this $g$ modulation for a while to generate a deep dip at $k_{\rm peak}$ and then turn off the modulation. The resulting $F_k$ with a deep dip at $k_{\rm peak}$ can be used as an initial state for the time evolution. To avoid other excitations, we need to tune the amplitude and the duration of the $g$ modulation, and reduce the amplitude slowly. In practice, the duration of the modulation should be long enough to make a deep dip at $k_{\rm peak}$, and the amplitude of the modulation has to be small enough to avoid exciting other modes. We find that this method can also generate Higgs-mode oscillations with a relatively long lifetime. However, because we cannot completely avoid other excitations, the oscillation decays in time, unlike in the first scheme of removing  particles directly. Figure~\ref{fig:g_modulation_at_peak} shows an example of the result from this method. There, we perform $g$ modulation with a constant amplitude from $t=0$ to $1000\,\hbar/E_F$, and then reduce the modulation amplitude to zero linearly from $t=1000\,\hbar/E_F$ to $1500\,\hbar/E_F$. Then, the system undergoes free evolution from $t=1500 \hbar/E_F$. At $t=1500\,\hbar/E_F$ just after removing the modulation, $F_k$ has a relatively deep dip around $k_{\rm peak}$ [Fig.~\ref{fig:g_modulation_at_peak}(a)], and then $|\Delta(t)|$ shows a quite stable oscillation from $t=1500\,\hbar/E_F$ [Fig.~\ref{fig:g_modulation_at_peak}(b)]. However, it is clearly discernible from Fig.~\ref{fig:g_modulation_at_peak}(b) that the amplitude of the oscillation  decreases, albeit slowly, in time (until it becomes irregular after very long time, which is not shown in the figure), unlike the results from direct removal of  particles [see, e.g., Figs.~\ref{fig:dn0_05_oscillation}(b)--\ref{fig:dn0_05_oscillation}(d)].

In summary, we have proposed a scheme to generate long-lived Higgs-mode oscillations in a homogeneous superfluid Fermi gas at zero temperature. First, we discussed the excitation of the Higgs mode by time-periodic modulation of the scattering length. Although Higgs-mode oscillations can be produced by this method, higher-harmonic excitations are also inevitably generated by the ``parametric up-conversion'' due to nonlinearity in the superfluid phase. Then, we proposed an alternative scheme to excite the Higgs mode by removing  particles around a certain absolute value of  momentum $\hbar k_{\rm hole}$. The removal of  particles can also be regarded as an unconventional type of quench. This scheme can generate very stable Higgs-mode oscillations with much longer lifetime and larger amplitude compared with the recent experiment~\cite{Behrle2018natphys}. We have demonstrated that persistent Higgs-mode oscillations with almost constant amplitude can be produced (see Fig.~\ref{fig:dn0_05_oscillation}). These persistent oscillations can be realized by tuning $k_{\rm hole}$ to  $k_{\rm peak}$, where $k_{\rm peak}$ is the wave number corresponding to the peak of the pair wave function $F_k$ in equilibrium. Furthermore, the amplitude of the Higgs-mode oscillation can be controlled by the number of removed particles. At the end, we proposed two ways to realize particle removal in experiment. It should be mentioned that our discussions are based on the uniform system. Nevertheless, given that preparing a quasihomogeneous system using a box potential is a fairly mature experimental technique \cite{Gaunt2013PRL}, our scheme can be readily implemented experimentally. We leave the case of a harmonic trapping potential to future work.

\begin{acknowledgments}
We thank B. Prasanna Venkatesh for helpful discussions and comments. G.W. was supported by the NSF of China (Grants No.~11975199 and No.~11674283), by the Zhejiang Provincial Natural Science Foundation Key Project (Grant No.~LZ19A050001), and by the Zhejiang University 100 Plan. Q.C. was supported by the NSF of China (Grant No. 11774309) and the Innovation Program
 for Quantum Science and Technology (Grant No. 2021ZD0301904). K.-T.X. was supported by the MOST of China (Grant No. G2022181023L) and NUAA (Grant No. YAT22005).

\medskip
G.L. and K.-T.X. contributed equally to this work.
\end{acknowledgments}

\bibliography{Higgsmode_refs_submit2}

\begin{thebibliography}{69}%
\makeatletter
\providecommand \@ifxundefined [1]{%
 \@ifx{#1\undefined}
}%
\providecommand \@ifnum [1]{%
 \ifnum #1\expandafter \@firstoftwo
 \else \expandafter \@secondoftwo
 \fi
}%
\providecommand \@ifx [1]{%
 \ifx #1\expandafter \@firstoftwo
 \else \expandafter \@secondoftwo
 \fi
}%
\providecommand \natexlab [1]{#1}%
\providecommand \enquote  [1]{``#1''}%
\providecommand \bibnamefont  [1]{#1}%
\providecommand \bibfnamefont [1]{#1}%
\providecommand \citenamefont [1]{#1}%
\providecommand \href@noop [0]{\@secondoftwo}%
\providecommand \href [0]{\begingroup \@sanitize@url \@href}%
\providecommand \@href[1]{\@@startlink{#1}\@@href}%
\providecommand \@@href[1]{\endgroup#1\@@endlink}%
\providecommand \@sanitize@url [0]{\catcode `\\12\catcode `\$12\catcode
  `\&12\catcode `\#12\catcode `\^12\catcode `\_12\catcode `\%12\relax}%
\providecommand \@@startlink[1]{}%
\providecommand \@@endlink[0]{}%
\providecommand \url  [0]{\begingroup\@sanitize@url \@url }%
\providecommand \@url [1]{\endgroup\@href {#1}{\urlprefix }}%
\providecommand \urlprefix  [0]{URL }%
\providecommand \Eprint [0]{\href }%
\providecommand \doibase [0]{https://doi.org/}%
\providecommand \selectlanguage [0]{\@gobble}%
\providecommand \bibinfo  [0]{\@secondoftwo}%
\providecommand \bibfield  [0]{\@secondoftwo}%
\providecommand \translation [1]{[#1]}%
\providecommand \BibitemOpen [0]{}%
\providecommand \bibitemStop [0]{}%
\providecommand \bibitemNoStop [0]{.\EOS\space}%
\providecommand \EOS [0]{\spacefactor3000\relax}%
\providecommand \BibitemShut  [1]{\csname bibitem#1\endcsname}%
\let\auto@bib@innerbib\@empty
\bibitem [{\citenamefont {Englert}\ and\ \citenamefont
  {Brout}(1964)}]{Englert1964PRL}%
  \BibitemOpen
  \bibfield  {author} {\bibinfo {author} {\bibfnamefont {F.}~\bibnamefont
  {Englert}}\ and\ \bibinfo {author} {\bibfnamefont {R.}~\bibnamefont
  {Brout}},\ }\bibfield  {title} {\bibinfo {title} {Broken {S}ymmetry and the
  {M}ass of {G}auge {V}ector {M}esons},\ }\href
  {https://doi.org/10.1103/PhysRevLett.13.321} {\bibfield  {journal} {\bibinfo
  {journal} {Phys. Rev. Lett.}\ }\textbf {\bibinfo {volume} {13}},\ \bibinfo
  {pages} {321} (\bibinfo {year} {1964})}\BibitemShut {NoStop}%
\bibitem [{\citenamefont {Higgs}(1964)}]{Higgs1964PRL}%
  \BibitemOpen
  \bibfield  {author} {\bibinfo {author} {\bibfnamefont {P.~W.}\ \bibnamefont
  {Higgs}},\ }\bibfield  {title} {\bibinfo {title} {Broken {S}ymmetries and the
  {M}asses of {G}auge {B}osons},\ }\href
  {https://doi.org/10.1103/PhysRevLett.13.508} {\bibfield  {journal} {\bibinfo
  {journal} {Phys. Rev. Lett.}\ }\textbf {\bibinfo {volume} {13}},\ \bibinfo
  {pages} {508} (\bibinfo {year} {1964})}\BibitemShut {NoStop}%
\bibitem [{\citenamefont {Guralnik}\ \emph {et~al.}(1964)\citenamefont
  {Guralnik}, \citenamefont {Hagen},\ and\ \citenamefont
  {Kibble}}]{Guralnik1964PRL}%
  \BibitemOpen
  \bibfield  {author} {\bibinfo {author} {\bibfnamefont {G.~S.}\ \bibnamefont
  {Guralnik}}, \bibinfo {author} {\bibfnamefont {C.~R.}\ \bibnamefont
  {Hagen}},\ and\ \bibinfo {author} {\bibfnamefont {T.~W.~B.}\ \bibnamefont
  {Kibble}},\ }\bibfield  {title} {\bibinfo {title} {Global {C}onservation
  {L}aws and {M}assless {P}articles},\ }\href
  {https://doi.org/10.1103/PhysRevLett.13.585} {\bibfield  {journal} {\bibinfo
  {journal} {Phys. Rev. Lett.}\ }\textbf {\bibinfo {volume} {13}},\ \bibinfo
  {pages} {585} (\bibinfo {year} {1964})}\BibitemShut {NoStop}%
\bibitem [{\citenamefont {Pekker}\ and\ \citenamefont
  {Varma}(2015)}]{Pekker2015AnnuRevCMP}%
  \BibitemOpen
  \bibfield  {author} {\bibinfo {author} {\bibfnamefont {D.}~\bibnamefont
  {Pekker}}\ and\ \bibinfo {author} {\bibfnamefont {C.~M.}\ \bibnamefont
  {Varma}},\ }\bibfield  {title} {\bibinfo {title} {{Amplitude/Higgs Modes in
  Condensed Matter Physics}},\ }\href
  {https://doi.org/10.1146/annurev-conmatphys-031214-014350} {\bibfield
  {journal} {\bibinfo  {journal} {Annu. Rev. Condens. Matter Phys.}\ }\textbf
  {\bibinfo {volume} {6}},\ \bibinfo {pages} {269} (\bibinfo {year}
  {2015})}\BibitemShut {NoStop}%
\bibitem [{\citenamefont {Shimano}\ and\ \citenamefont
  {Tsuji}(2020)}]{Shimano2020AnnuRevCMP}%
  \BibitemOpen
  \bibfield  {author} {\bibinfo {author} {\bibfnamefont {R.}~\bibnamefont
  {Shimano}}\ and\ \bibinfo {author} {\bibfnamefont {N.}~\bibnamefont
  {Tsuji}},\ }\bibfield  {title} {\bibinfo {title} {{Higgs Mode in
  Superconductors}},\ }\href
  {https://doi.org/10.1146/annurev-conmatphys-031119-050813} {\bibfield
  {journal} {\bibinfo  {journal} {Annu. Rev. Condens. Matter Phys.}\ }\textbf
  {\bibinfo {volume} {11}},\ \bibinfo {pages} {103} (\bibinfo {year}
  {2020})}\BibitemShut {NoStop}%
\bibitem [{\citenamefont {Volovik}\ and\ \citenamefont
  {Zubkov}(2014)}]{Volovik2014JLTP}%
  \BibitemOpen
  \bibfield  {author} {\bibinfo {author} {\bibfnamefont {G.~E.}\ \bibnamefont
  {Volovik}}\ and\ \bibinfo {author} {\bibfnamefont {M.~A.}\ \bibnamefont
  {Zubkov}},\ }\bibfield  {title} {\bibinfo {title} {{Higgs Bosons in Particle
  Physics and in Condensed Matter}},\ }\href
  {https://doi.org/10.1007/s10909-013-0905-7} {\bibfield  {journal} {\bibinfo
  {journal} {Journal of Low Temperature Physics}\ }\textbf {\bibinfo {volume}
  {175}},\ \bibinfo {pages} {486} (\bibinfo {year} {2014})}\BibitemShut
  {NoStop}%
\bibitem [{\citenamefont {Sooryakumar}\ and\ \citenamefont
  {Klein}(1980)}]{Sooryakumar1980PRL}%
  \BibitemOpen
  \bibfield  {author} {\bibinfo {author} {\bibfnamefont {R.}~\bibnamefont
  {Sooryakumar}}\ and\ \bibinfo {author} {\bibfnamefont {M.~V.}\ \bibnamefont
  {Klein}},\ }\bibfield  {title} {\bibinfo {title} {{Raman Scattering by
  Superconducting-Gap Excitations and Their Coupling to Charge-Density
  Waves}},\ }\href {https://doi.org/10.1103/PhysRevLett.45.660} {\bibfield
  {journal} {\bibinfo  {journal} {Phys. Rev. Lett.}\ }\textbf {\bibinfo
  {volume} {45}},\ \bibinfo {pages} {660} (\bibinfo {year} {1980})}\BibitemShut
  {NoStop}%
\bibitem [{\citenamefont {Bissbort}\ \emph {et~al.}(2011)\citenamefont
  {Bissbort}, \citenamefont {G\"otze}, \citenamefont {Li}, \citenamefont
  {Heinze}, \citenamefont {Krauser}, \citenamefont {Weinberg}, \citenamefont
  {Becker}, \citenamefont {Sengstock},\ and\ \citenamefont
  {Hofstetter}}]{Bissbort2011PRL}%
  \BibitemOpen
  \bibfield  {author} {\bibinfo {author} {\bibfnamefont {U.}~\bibnamefont
  {Bissbort}}, \bibinfo {author} {\bibfnamefont {S.}~\bibnamefont {G\"otze}},
  \bibinfo {author} {\bibfnamefont {Y.}~\bibnamefont {Li}}, \bibinfo {author}
  {\bibfnamefont {J.}~\bibnamefont {Heinze}}, \bibinfo {author} {\bibfnamefont
  {J.~S.}\ \bibnamefont {Krauser}}, \bibinfo {author} {\bibfnamefont
  {M.}~\bibnamefont {Weinberg}}, \bibinfo {author} {\bibfnamefont
  {C.}~\bibnamefont {Becker}}, \bibinfo {author} {\bibfnamefont
  {K.}~\bibnamefont {Sengstock}},\ and\ \bibinfo {author} {\bibfnamefont
  {W.}~\bibnamefont {Hofstetter}},\ }\bibfield  {title} {\bibinfo {title}
  {{Detecting the Amplitude Mode of Strongly Interacting Lattice Bosons by
  Bragg Scattering}},\ }\href {https://doi.org/10.1103/PhysRevLett.106.205303}
  {\bibfield  {journal} {\bibinfo  {journal} {Phys. Rev. Lett.}\ }\textbf
  {\bibinfo {volume} {106}},\ \bibinfo {pages} {205303} (\bibinfo {year}
  {2011})}\BibitemShut {NoStop}%
\bibitem [{\citenamefont {Endres}\ \emph {et~al.}(2012)\citenamefont {Endres},
  \citenamefont {Fukuhara}, \citenamefont {Pekker}, \citenamefont {Cheneau},
  \citenamefont {Schau\ss}, \citenamefont {Gross}, \citenamefont {Demler},
  \citenamefont {Kuhr},\ and\ \citenamefont {Bloch}}]{Endres2012Nature}%
  \BibitemOpen
  \bibfield  {author} {\bibinfo {author} {\bibfnamefont {M.}~\bibnamefont
  {Endres}}, \bibinfo {author} {\bibfnamefont {T.}~\bibnamefont {Fukuhara}},
  \bibinfo {author} {\bibfnamefont {D.}~\bibnamefont {Pekker}}, \bibinfo
  {author} {\bibfnamefont {M.}~\bibnamefont {Cheneau}}, \bibinfo {author}
  {\bibfnamefont {P.}~\bibnamefont {Schau\ss}}, \bibinfo {author}
  {\bibfnamefont {C.}~\bibnamefont {Gross}}, \bibinfo {author} {\bibfnamefont
  {E.}~\bibnamefont {Demler}}, \bibinfo {author} {\bibfnamefont
  {S.}~\bibnamefont {Kuhr}},\ and\ \bibinfo {author} {\bibfnamefont
  {I.}~\bibnamefont {Bloch}},\ }\bibfield  {title} {\bibinfo {title} {{The
  `Higgs' Amplitude Mode at the Two-Dimensional Superfluid/Mott Insulator
  Transition}},\ }\href {https://doi.org/10.1038/nature11255} {\bibfield
  {journal} {\bibinfo  {journal} {Nature (London)}\ }\textbf {\bibinfo {volume}
  {487}},\ \bibinfo {pages} {454} (\bibinfo {year} {2012})}\BibitemShut
  {NoStop}%
\bibitem [{\citenamefont {Matsunaga}\ \emph {et~al.}(2013)\citenamefont
  {Matsunaga}, \citenamefont {Hamada}, \citenamefont {Makise}, \citenamefont
  {Uzawa}, \citenamefont {Terai}, \citenamefont {Wang},\ and\ \citenamefont
  {Shimano}}]{Matsunaga2013PRL}%
  \BibitemOpen
  \bibfield  {author} {\bibinfo {author} {\bibfnamefont {R.}~\bibnamefont
  {Matsunaga}}, \bibinfo {author} {\bibfnamefont {Y.~I.}\ \bibnamefont
  {Hamada}}, \bibinfo {author} {\bibfnamefont {K.}~\bibnamefont {Makise}},
  \bibinfo {author} {\bibfnamefont {Y.}~\bibnamefont {Uzawa}}, \bibinfo
  {author} {\bibfnamefont {H.}~\bibnamefont {Terai}}, \bibinfo {author}
  {\bibfnamefont {Z.}~\bibnamefont {Wang}},\ and\ \bibinfo {author}
  {\bibfnamefont {R.}~\bibnamefont {Shimano}},\ }\bibfield  {title} {\bibinfo
  {title} {{Higgs Amplitude Mode in the BCS Superconductors
  \text{${\mathrm{Nb}}_{1\mathrm{\text{\ensuremath{-}}}x}{\mathrm{Ti}}_{x}\mathbf{N}$}
  Induced by Terahertz Pulse Excitation}},\ }\href
  {https://doi.org/10.1103/PhysRevLett.111.057002} {\bibfield  {journal}
  {\bibinfo  {journal} {Phys. Rev. Lett.}\ }\textbf {\bibinfo {volume} {111}},\
  \bibinfo {pages} {057002} (\bibinfo {year} {2013})}\BibitemShut {NoStop}%
\bibitem [{\citenamefont {Mansart}\ \emph {et~al.}(2013)\citenamefont
  {Mansart}, \citenamefont {Lorenzana}, \citenamefont {Mann}, \citenamefont
  {Odeh}, \citenamefont {Scarongella}, \citenamefont {Chergui},\ and\
  \citenamefont {Carbone}}]{Mansart2013PNAS}%
  \BibitemOpen
  \bibfield  {author} {\bibinfo {author} {\bibfnamefont {B.}~\bibnamefont
  {Mansart}}, \bibinfo {author} {\bibfnamefont {J.}~\bibnamefont {Lorenzana}},
  \bibinfo {author} {\bibfnamefont {A.}~\bibnamefont {Mann}}, \bibinfo {author}
  {\bibfnamefont {A.}~\bibnamefont {Odeh}}, \bibinfo {author} {\bibfnamefont
  {M.}~\bibnamefont {Scarongella}}, \bibinfo {author} {\bibfnamefont
  {M.}~\bibnamefont {Chergui}},\ and\ \bibinfo {author} {\bibfnamefont
  {F.}~\bibnamefont {Carbone}},\ }\bibfield  {title} {\bibinfo {title}
  {{Coupling of a High-Energy Excitation to Superconducting Quasiparticles in a
  Cuprate from Coherent Charge Fluctuation Spectroscopy}},\ }\href
  {https://doi.org/10.1073/pnas.1218742110} {\bibfield  {journal} {\bibinfo
  {journal} {Proc. Natl. Acad. Sci. USA}\ }\textbf {\bibinfo {volume} {110}},\
  \bibinfo {pages} {4539} (\bibinfo {year} {2013})}\BibitemShut {NoStop}%
\bibitem [{\citenamefont {Sherman}\ \emph {et~al.}(2015)\citenamefont
  {Sherman}, \citenamefont {Pracht}, \citenamefont {Gorshunov}, \citenamefont
  {Poran}, \citenamefont {Jesudasan}, \citenamefont {Chand}, \citenamefont
  {Raychaudhuri}, \citenamefont {Swanson}, \citenamefont {Trivedi},
  \citenamefont {Auerbach}, \citenamefont {Scheffler}, \citenamefont
  {Frydman},\ and\ \citenamefont {Dressel}}]{Sherman2015NatPhys}%
  \BibitemOpen
  \bibfield  {author} {\bibinfo {author} {\bibfnamefont {D.}~\bibnamefont
  {Sherman}}, \bibinfo {author} {\bibfnamefont {U.~S.}\ \bibnamefont {Pracht}},
  \bibinfo {author} {\bibfnamefont {B.}~\bibnamefont {Gorshunov}}, \bibinfo
  {author} {\bibfnamefont {S.}~\bibnamefont {Poran}}, \bibinfo {author}
  {\bibfnamefont {J.}~\bibnamefont {Jesudasan}}, \bibinfo {author}
  {\bibfnamefont {M.}~\bibnamefont {Chand}}, \bibinfo {author} {\bibfnamefont
  {P.}~\bibnamefont {Raychaudhuri}}, \bibinfo {author} {\bibfnamefont
  {M.}~\bibnamefont {Swanson}}, \bibinfo {author} {\bibfnamefont
  {N.}~\bibnamefont {Trivedi}}, \bibinfo {author} {\bibfnamefont
  {A.}~\bibnamefont {Auerbach}}, \bibinfo {author} {\bibfnamefont
  {M.}~\bibnamefont {Scheffler}}, \bibinfo {author} {\bibfnamefont
  {A.}~\bibnamefont {Frydman}},\ and\ \bibinfo {author} {\bibfnamefont
  {M.}~\bibnamefont {Dressel}},\ }\bibfield  {title} {\bibinfo {title} {{The
  Higgs Mode in Disordered Superconductors Close to a Quantum Phase
  Transition}},\ }\href {https://doi.org/10.1038/nphys3227} {\bibfield
  {journal} {\bibinfo  {journal} {Nat. Phys.}\ }\textbf {\bibinfo {volume}
  {11}},\ \bibinfo {pages} {188} (\bibinfo {year} {2015})}\BibitemShut
  {NoStop}%
\bibitem [{\citenamefont {Hoang}\ \emph {et~al.}(2016)\citenamefont {Hoang},
  \citenamefont {Bharath}, \citenamefont {Boguslawski}, \citenamefont {Anquez},
  \citenamefont {Robbins},\ and\ \citenamefont {Chapman}}]{Hoang2016PNAS}%
  \BibitemOpen
  \bibfield  {author} {\bibinfo {author} {\bibfnamefont {T.~M.}\ \bibnamefont
  {Hoang}}, \bibinfo {author} {\bibfnamefont {H.~M.}\ \bibnamefont {Bharath}},
  \bibinfo {author} {\bibfnamefont {M.~J.}\ \bibnamefont {Boguslawski}},
  \bibinfo {author} {\bibfnamefont {M.}~\bibnamefont {Anquez}}, \bibinfo
  {author} {\bibfnamefont {B.~A.}\ \bibnamefont {Robbins}},\ and\ \bibinfo
  {author} {\bibfnamefont {M.~S.}\ \bibnamefont {Chapman}},\ }\bibfield
  {title} {\bibinfo {title} {{Adiabatic Quenches and Characterization of
  Amplitude Excitations in a Continuous Quantum Phase Transition}},\ }\href
  {https://www.pnas.org/doi/abs/10.1073/pnas.1600267113} {\bibfield  {journal}
  {\bibinfo  {journal} {Proc. Natl. Acad. Sci. USA}\ }\textbf {\bibinfo
  {volume} {113}},\ \bibinfo {pages} {9475} (\bibinfo {year}
  {2016})}\BibitemShut {NoStop}%
\bibitem [{\citenamefont {Jain}\ \emph {et~al.}(2017)\citenamefont {Jain},
  \citenamefont {Krautloher}, \citenamefont {Porras}, \citenamefont {Ryu},
  \citenamefont {Chen}, \citenamefont {Abernathy}, \citenamefont {Park},
  \citenamefont {Ivanov}, \citenamefont {Chaloupka}, \citenamefont
  {Khaliullin}, \citenamefont {Keimer},\ and\ \citenamefont
  {Kim}}]{Jain2017natphys}%
  \BibitemOpen
  \bibfield  {author} {\bibinfo {author} {\bibfnamefont {A.}~\bibnamefont
  {Jain}}, \bibinfo {author} {\bibfnamefont {M.}~\bibnamefont {Krautloher}},
  \bibinfo {author} {\bibfnamefont {J.}~\bibnamefont {Porras}}, \bibinfo
  {author} {\bibfnamefont {G.~H.}\ \bibnamefont {Ryu}}, \bibinfo {author}
  {\bibfnamefont {D.~P.}\ \bibnamefont {Chen}}, \bibinfo {author}
  {\bibfnamefont {D.~L.}\ \bibnamefont {Abernathy}}, \bibinfo {author}
  {\bibfnamefont {J.~T.}\ \bibnamefont {Park}}, \bibinfo {author}
  {\bibfnamefont {A.}~\bibnamefont {Ivanov}}, \bibinfo {author} {\bibfnamefont
  {J.}~\bibnamefont {Chaloupka}}, \bibinfo {author} {\bibfnamefont
  {G.}~\bibnamefont {Khaliullin}}, \bibinfo {author} {\bibfnamefont
  {B.}~\bibnamefont {Keimer}},\ and\ \bibinfo {author} {\bibfnamefont {B.~J.}\
  \bibnamefont {Kim}},\ }\bibfield  {title} {\bibinfo {title} {{Higgs Mode and
  its Decay in a Two-Dimensional Antiferromagnet}},\ }\href
  {https://doi.org/10.1038/nphys4077} {\bibfield  {journal} {\bibinfo
  {journal} {Nat. Phys.}\ }\textbf {\bibinfo {volume} {13}},\ \bibinfo {pages}
  {633} (\bibinfo {year} {2017})}\BibitemShut {NoStop}%
\bibitem [{\citenamefont {Hong}\ \emph {et~al.}(2017)\citenamefont {Hong},
  \citenamefont {Matsumoto}, \citenamefont {Qiu}, \citenamefont {Chen},
  \citenamefont {Gentile}, \citenamefont {Watson}, \citenamefont {Awwadi},
  \citenamefont {Turnbull}, \citenamefont {Dissanayake}, \citenamefont
  {Agrawal}, \citenamefont {Toft-Petersen}, \citenamefont {Klemke},
  \citenamefont {Coester}, \citenamefont {Schmidt},\ and\ \citenamefont
  {Tennant}}]{Hong2017NatPhys}%
  \BibitemOpen
  \bibfield  {author} {\bibinfo {author} {\bibfnamefont {T.}~\bibnamefont
  {Hong}}, \bibinfo {author} {\bibfnamefont {M.}~\bibnamefont {Matsumoto}},
  \bibinfo {author} {\bibfnamefont {Y.}~\bibnamefont {Qiu}}, \bibinfo {author}
  {\bibfnamefont {W.}~\bibnamefont {Chen}}, \bibinfo {author} {\bibfnamefont
  {T.~R.}\ \bibnamefont {Gentile}}, \bibinfo {author} {\bibfnamefont
  {S.}~\bibnamefont {Watson}}, \bibinfo {author} {\bibfnamefont {F.~F.}\
  \bibnamefont {Awwadi}}, \bibinfo {author} {\bibfnamefont {M.~M.}\
  \bibnamefont {Turnbull}}, \bibinfo {author} {\bibfnamefont {S.~E.}\
  \bibnamefont {Dissanayake}}, \bibinfo {author} {\bibfnamefont
  {H.}~\bibnamefont {Agrawal}}, \bibinfo {author} {\bibfnamefont
  {R.}~\bibnamefont {Toft-Petersen}}, \bibinfo {author} {\bibfnamefont
  {B.}~\bibnamefont {Klemke}}, \bibinfo {author} {\bibfnamefont
  {K.}~\bibnamefont {Coester}}, \bibinfo {author} {\bibfnamefont {K.~P.}\
  \bibnamefont {Schmidt}},\ and\ \bibinfo {author} {\bibfnamefont {D.~A.}\
  \bibnamefont {Tennant}},\ }\bibfield  {title} {\bibinfo {title} {{Higgs
  Amplitude Mode in a Two-Dimensional Quantum Antiferromagnet near the Quantum
  Critical Point}},\ }\href {https://doi.org/10.1038/nphys4182} {\bibfield
  {journal} {\bibinfo  {journal} {Nat. Phys.}\ }\textbf {\bibinfo {volume}
  {13}},\ \bibinfo {pages} {638} (\bibinfo {year} {2017})}\BibitemShut
  {NoStop}%
\bibitem [{\citenamefont {L{\'e}onard}\ \emph {et~al.}(2017)\citenamefont
  {L{\'e}onard}, \citenamefont {Morales}, \citenamefont {Zupancic},
  \citenamefont {Donner},\ and\ \citenamefont
  {Esslinger}}]{Leonard2017Science}%
  \BibitemOpen
  \bibfield  {author} {\bibinfo {author} {\bibfnamefont {J.}~\bibnamefont
  {L{\'e}onard}}, \bibinfo {author} {\bibfnamefont {A.}~\bibnamefont
  {Morales}}, \bibinfo {author} {\bibfnamefont {P.}~\bibnamefont {Zupancic}},
  \bibinfo {author} {\bibfnamefont {T.}~\bibnamefont {Donner}},\ and\ \bibinfo
  {author} {\bibfnamefont {T.}~\bibnamefont {Esslinger}},\ }\bibfield  {title}
  {\bibinfo {title} {{Monitoring and Manipulating Higgs and Goldstone Modes in
  a Supersolid Quantum Gas}},\ }\href {https://doi.org/10.1126/science.aan2608}
  {\bibfield  {journal} {\bibinfo  {journal} {Science}\ }\textbf {\bibinfo
  {volume} {358}},\ \bibinfo {pages} {1415} (\bibinfo {year}
  {2017})}\BibitemShut {NoStop}%
\bibitem [{\citenamefont {Behrle}\ \emph {et~al.}(2018)\citenamefont {Behrle},
  \citenamefont {Harrison}, \citenamefont {Kombe}, \citenamefont {Gao},
  \citenamefont {Link}, \citenamefont {Bernier}, \citenamefont {Kollath},\ and\
  \citenamefont {K{\"o}hl}}]{Behrle2018natphys}%
  \BibitemOpen
  \bibfield  {author} {\bibinfo {author} {\bibfnamefont {A.}~\bibnamefont
  {Behrle}}, \bibinfo {author} {\bibfnamefont {T.}~\bibnamefont {Harrison}},
  \bibinfo {author} {\bibfnamefont {J.}~\bibnamefont {Kombe}}, \bibinfo
  {author} {\bibfnamefont {K.}~\bibnamefont {Gao}}, \bibinfo {author}
  {\bibfnamefont {M.}~\bibnamefont {Link}}, \bibinfo {author} {\bibfnamefont
  {J.-S.}\ \bibnamefont {Bernier}}, \bibinfo {author} {\bibfnamefont
  {C.}~\bibnamefont {Kollath}},\ and\ \bibinfo {author} {\bibfnamefont
  {M.}~\bibnamefont {K{\"o}hl}},\ }\bibfield  {title} {\bibinfo {title} {{Higgs
  Mode in a Strongly Interacting Fermionic Superfluid}},\ }\href
  {https://doi.org/10.1038/s41567-018-0128-6} {\bibfield  {journal} {\bibinfo
  {journal} {Nat. Phys.}\ }\textbf {\bibinfo {volume} {14}},\ \bibinfo {pages}
  {781} (\bibinfo {year} {2018})}\BibitemShut {NoStop}%
\bibitem [{\citenamefont {\text{ATLAS Collaboration}}(2012)}]{Aad2012PLB}%
  \BibitemOpen
  \bibfield  {author} {\bibinfo {author} {\bibnamefont {\text{ATLAS
  Collaboration}}},\ }\bibfield  {title} {\bibinfo {title} {{Observation of a
  New Particle in the Search for the Standard Model Higgs Boson with the ATLAS
  Detector at the LHC}},\ }\href
  {https://doi.org/https://doi.org/10.1016/j.physletb.2012.08.020} {\bibfield
  {journal} {\bibinfo  {journal} {Phys. Lett. B}\ }\textbf {\bibinfo {volume}
  {716}},\ \bibinfo {pages} {1} (\bibinfo {year} {2012})}\BibitemShut {NoStop}%
\bibitem [{\citenamefont {\text{CMS Collaboration}}(2012)}]{Chatrchyan2012PLB}%
  \BibitemOpen
  \bibfield  {author} {\bibinfo {author} {\bibnamefont {\text{CMS
  Collaboration}}},\ }\bibfield  {title} {\bibinfo {title} {{Observation of a
  New Boson at a Mass of 125 GeV with the CMS Experiment at the LHC}},\ }\href
  {https://doi.org/https://doi.org/10.1016/j.physletb.2012.08.021} {\bibfield
  {journal} {\bibinfo  {journal} {Phys. Lett. B}\ }\textbf {\bibinfo {volume}
  {716}},\ \bibinfo {pages} {30} (\bibinfo {year} {2012})}\BibitemShut
  {NoStop}%
\bibitem [{\citenamefont {Volkov}\ and\ \citenamefont
  {Kogan}(1973)}]{Volkov1973JETP}%
  \BibitemOpen
  \bibfield  {author} {\bibinfo {author} {\bibfnamefont {A.~F.}\ \bibnamefont
  {Volkov}}\ and\ \bibinfo {author} {\bibfnamefont {S.~M.}\ \bibnamefont
  {Kogan}},\ }\bibfield  {title} {\bibinfo {title} {{Collisionless Relaxation
  of the Energy Gap in Superconductors}},\ }\href
  {http://www.jetp.ras.ru/cgi-bin/e/index/r/65/5/p2038?a=list} {\bibfield
  {journal} {\bibinfo  {journal} {Zh. Eksp. Teor. Fiz.}\ }\textbf {\bibinfo
  {volume} {65}},\ \bibinfo {pages} {2038} (\bibinfo {year} {1973})},\ \bibinfo
  {note} {[Sov. Phys. JETP {\bf 38}, 1018 (1974)]}\BibitemShut {NoStop}%
\bibitem [{\citenamefont {Gurarie}(2009)}]{Gurarie2009PRL}%
  \BibitemOpen
  \bibfield  {author} {\bibinfo {author} {\bibfnamefont {V.}~\bibnamefont
  {Gurarie}},\ }\bibfield  {title} {\bibinfo {title} {{Nonequilibrium Dynamics
  of Weakly and Strongly Paired Superconductors}},\ }\href
  {https://doi.org/10.1103/PhysRevLett.103.075301} {\bibfield  {journal}
  {\bibinfo  {journal} {Phys. Rev. Lett.}\ }\textbf {\bibinfo {volume} {103}},\
  \bibinfo {pages} {075301} (\bibinfo {year} {2009})}\BibitemShut {NoStop}%
\bibitem [{\citenamefont {Littlewood}\ and\ \citenamefont
  {Varma}(1981)}]{Littlewood1981PRL}%
  \BibitemOpen
  \bibfield  {author} {\bibinfo {author} {\bibfnamefont {P.~B.}\ \bibnamefont
  {Littlewood}}\ and\ \bibinfo {author} {\bibfnamefont {C.~M.}\ \bibnamefont
  {Varma}},\ }\bibfield  {title} {\bibinfo {title} {{Gauge-Invariant Theory of
  the Dynamical Interaction of Charge Density Waves and Superconductivity}},\
  }\href {https://doi.org/10.1103/PhysRevLett.47.811} {\bibfield  {journal}
  {\bibinfo  {journal} {Phys. Rev. Lett.}\ }\textbf {\bibinfo {volume} {47}},\
  \bibinfo {pages} {811} (\bibinfo {year} {1981})}\BibitemShut {NoStop}%
\bibitem [{\citenamefont {Littlewood}\ and\ \citenamefont
  {Varma}(1982)}]{Littlewood1982PRB}%
  \BibitemOpen
  \bibfield  {author} {\bibinfo {author} {\bibfnamefont {P.~B.}\ \bibnamefont
  {Littlewood}}\ and\ \bibinfo {author} {\bibfnamefont {C.~M.}\ \bibnamefont
  {Varma}},\ }\bibfield  {title} {\bibinfo {title} {{Amplitude Collective Modes
  in Superconductors and Their Coupling to Charge-Density Waves}},\ }\href
  {https://doi.org/10.1103/PhysRevB.26.4883} {\bibfield  {journal} {\bibinfo
  {journal} {Phys. Rev. B}\ }\textbf {\bibinfo {volume} {26}},\ \bibinfo
  {pages} {4883} (\bibinfo {year} {1982})}\BibitemShut {NoStop}%
\bibitem [{\citenamefont {DeMarco}\ and\ \citenamefont
  {Jin}(1999)}]{Demarco1999science}%
  \BibitemOpen
  \bibfield  {author} {\bibinfo {author} {\bibfnamefont {B.}~\bibnamefont
  {DeMarco}}\ and\ \bibinfo {author} {\bibfnamefont {D.~S.}\ \bibnamefont
  {Jin}},\ }\bibfield  {title} {\bibinfo {title} {{Onset of Fermi Degeneracy in
  a Trapped Atomic Gas}},\ }\href
  {https://doi.org/10.1126/science.285.5434.1703} {\bibfield  {journal}
  {\bibinfo  {journal} {Science}\ }\textbf {\bibinfo {volume} {285}},\ \bibinfo
  {pages} {1703} (\bibinfo {year} {1999})}\BibitemShut {NoStop}%
\bibitem [{\citenamefont {Andreev}\ \emph {et~al.}(2004)\citenamefont
  {Andreev}, \citenamefont {Gurarie},\ and\ \citenamefont
  {Radzihovsky}}]{Andreev2004PRL}%
  \BibitemOpen
  \bibfield  {author} {\bibinfo {author} {\bibfnamefont {A.~V.}\ \bibnamefont
  {Andreev}}, \bibinfo {author} {\bibfnamefont {V.}~\bibnamefont {Gurarie}},\
  and\ \bibinfo {author} {\bibfnamefont {L.}~\bibnamefont {Radzihovsky}},\
  }\bibfield  {title} {\bibinfo {title} {{Nonequilibrium Dynamics and
  Thermodynamics of a Degenerate Fermi Gas Across a Feshbach Resonance}},\
  }\href {https://doi.org/10.1103/PhysRevLett.93.130402} {\bibfield  {journal}
  {\bibinfo  {journal} {Phys. Rev. Lett.}\ }\textbf {\bibinfo {volume} {93}},\
  \bibinfo {pages} {130402} (\bibinfo {year} {2004})}\BibitemShut {NoStop}%
\bibitem [{\citenamefont {Barankov}\ \emph {et~al.}(2004)\citenamefont
  {Barankov}, \citenamefont {Levitov},\ and\ \citenamefont
  {Spivak}}]{Barankov2004PRL}%
  \BibitemOpen
  \bibfield  {author} {\bibinfo {author} {\bibfnamefont {R.~A.}\ \bibnamefont
  {Barankov}}, \bibinfo {author} {\bibfnamefont {L.~S.}\ \bibnamefont
  {Levitov}},\ and\ \bibinfo {author} {\bibfnamefont {B.~Z.}\ \bibnamefont
  {Spivak}},\ }\bibfield  {title} {\bibinfo {title} {{Collective Rabi
  Oscillations and Solitons in a Time-Dependent BCS Pairing Problem}},\ }\href
  {https://doi.org/10.1103/PhysRevLett.93.160401} {\bibfield  {journal}
  {\bibinfo  {journal} {Phys. Rev. Lett.}\ }\textbf {\bibinfo {volume} {93}},\
  \bibinfo {pages} {160401} (\bibinfo {year} {2004})}\BibitemShut {NoStop}%
\bibitem [{\citenamefont {Szyma\ifmmode~\acute{n}\else \'{n}\fi{}ska}\ \emph
  {et~al.}(2005)\citenamefont {Szyma\ifmmode~\acute{n}\else \'{n}\fi{}ska},
  \citenamefont {Simons},\ and\ \citenamefont {Burnett}}]{Szymanska2005PRL}%
  \BibitemOpen
  \bibfield  {author} {\bibinfo {author} {\bibfnamefont {M.~H.}\ \bibnamefont
  {Szyma\ifmmode~\acute{n}\else \'{n}\fi{}ska}}, \bibinfo {author}
  {\bibfnamefont {B.~D.}\ \bibnamefont {Simons}},\ and\ \bibinfo {author}
  {\bibfnamefont {K.}~\bibnamefont {Burnett}},\ }\bibfield  {title} {\bibinfo
  {title} {{Dynamics of the BCS-BEC Crossover in a Degenerate Fermi Gas}},\
  }\href {https://doi.org/10.1103/PhysRevLett.94.170402} {\bibfield  {journal}
  {\bibinfo  {journal} {Phys. Rev. Lett.}\ }\textbf {\bibinfo {volume} {94}},\
  \bibinfo {pages} {170402} (\bibinfo {year} {2005})}\BibitemShut {NoStop}%
\bibitem [{\citenamefont {Yuzbashyan}\ \emph {et~al.}(2006)\citenamefont
  {Yuzbashyan}, \citenamefont {Tsyplyatyev},\ and\ \citenamefont
  {Altshuler}}]{Yuzbashyan2006PRL}%
  \BibitemOpen
  \bibfield  {author} {\bibinfo {author} {\bibfnamefont {E.~A.}\ \bibnamefont
  {Yuzbashyan}}, \bibinfo {author} {\bibfnamefont {O.}~\bibnamefont
  {Tsyplyatyev}},\ and\ \bibinfo {author} {\bibfnamefont {B.~L.}\ \bibnamefont
  {Altshuler}},\ }\bibfield  {title} {\bibinfo {title} {{Relaxation and
  Persistent Oscillations of the Order Parameter in Fermionic Condensates}},\
  }\href {https://doi.org/10.1103/PhysRevLett.96.097005} {\bibfield  {journal}
  {\bibinfo  {journal} {Phys. Rev. Lett.}\ }\textbf {\bibinfo {volume} {96}},\
  \bibinfo {pages} {097005} (\bibinfo {year} {2006})}\BibitemShut {NoStop}%
\bibitem [{\citenamefont {Barankov}\ and\ \citenamefont
  {Levitov}(2006)}]{Barankov2006PRL}%
  \BibitemOpen
  \bibfield  {author} {\bibinfo {author} {\bibfnamefont {R.~A.}\ \bibnamefont
  {Barankov}}\ and\ \bibinfo {author} {\bibfnamefont {L.~S.}\ \bibnamefont
  {Levitov}},\ }\bibfield  {title} {\bibinfo {title} {{Synchronization in the
  BCS Pairing Dynamics as a Critical Phenomenon}},\ }\href
  {https://doi.org/10.1103/PhysRevLett.96.230403} {\bibfield  {journal}
  {\bibinfo  {journal} {Phys. Rev. Lett.}\ }\textbf {\bibinfo {volume} {96}},\
  \bibinfo {pages} {230403} (\bibinfo {year} {2006})}\BibitemShut {NoStop}%
\bibitem [{\citenamefont {Bulgac}\ and\ \citenamefont
  {Yoon}(2009)}]{Bulgac2009PRL}%
  \BibitemOpen
  \bibfield  {author} {\bibinfo {author} {\bibfnamefont {A.}~\bibnamefont
  {Bulgac}}\ and\ \bibinfo {author} {\bibfnamefont {S.}~\bibnamefont {Yoon}},\
  }\bibfield  {title} {\bibinfo {title} {{Large Amplitude Dynamics of the
  Pairing Correlations in a Unitary Fermi Gas}},\ }\href
  {https://doi.org/10.1103/PhysRevLett.102.085302} {\bibfield  {journal}
  {\bibinfo  {journal} {Phys. Rev. Lett.}\ }\textbf {\bibinfo {volume} {102}},\
  \bibinfo {pages} {085302} (\bibinfo {year} {2009})}\BibitemShut {NoStop}%
\bibitem [{\citenamefont {Hannibal}\ \emph {et~al.}(2015)\citenamefont
  {Hannibal}, \citenamefont {Kettmann}, \citenamefont {Croitoru}, \citenamefont
  {Vagov}, \citenamefont {Axt},\ and\ \citenamefont {Kuhn}}]{Hannibal2015PRA}%
  \BibitemOpen
  \bibfield  {author} {\bibinfo {author} {\bibfnamefont {S.}~\bibnamefont
  {Hannibal}}, \bibinfo {author} {\bibfnamefont {P.}~\bibnamefont {Kettmann}},
  \bibinfo {author} {\bibfnamefont {M.~D.}\ \bibnamefont {Croitoru}}, \bibinfo
  {author} {\bibfnamefont {A.}~\bibnamefont {Vagov}}, \bibinfo {author}
  {\bibfnamefont {V.~M.}\ \bibnamefont {Axt}},\ and\ \bibinfo {author}
  {\bibfnamefont {T.}~\bibnamefont {Kuhn}},\ }\bibfield  {title} {\bibinfo
  {title} {{Quench Dynamics of an Ultracold Fermi Gas in the BCS Regime:
  Spectral Properties and Confinement-Induced Breakdown of the Higgs Mode}},\
  }\href {https://doi.org/10.1103/PhysRevA.91.043630} {\bibfield  {journal}
  {\bibinfo  {journal} {Phys. Rev. A}\ }\textbf {\bibinfo {volume} {91}},\
  \bibinfo {pages} {043630} (\bibinfo {year} {2015})}\BibitemShut {NoStop}%
\bibitem [{\citenamefont {Yoon}\ and\ \citenamefont
  {Watanabe}(2017)}]{Yoon2017PRL}%
  \BibitemOpen
  \bibfield  {author} {\bibinfo {author} {\bibfnamefont {S.}~\bibnamefont
  {Yoon}}\ and\ \bibinfo {author} {\bibfnamefont {G.}~\bibnamefont
  {Watanabe}},\ }\bibfield  {title} {\bibinfo {title} {{Pairing Dynamics of
  Polar States in a Quenched $p$-Wave Superfluid Fermi Gas}},\ }\href
  {https://doi.org/10.1103/PhysRevLett.119.100401} {\bibfield  {journal}
  {\bibinfo  {journal} {Phys. Rev. Lett.}\ }\textbf {\bibinfo {volume} {119}},\
  \bibinfo {pages} {100401} (\bibinfo {year} {2017})}\BibitemShut {NoStop}%
\bibitem [{\citenamefont {Hannibal}\ \emph {et~al.}(2018)\citenamefont
  {Hannibal}, \citenamefont {Kettmann}, \citenamefont {Croitoru}, \citenamefont
  {Axt},\ and\ \citenamefont {Kuhn}}]{Hannibal2018PRA}%
  \BibitemOpen
  \bibfield  {author} {\bibinfo {author} {\bibfnamefont {S.}~\bibnamefont
  {Hannibal}}, \bibinfo {author} {\bibfnamefont {P.}~\bibnamefont {Kettmann}},
  \bibinfo {author} {\bibfnamefont {M.~D.}\ \bibnamefont {Croitoru}}, \bibinfo
  {author} {\bibfnamefont {V.~M.}\ \bibnamefont {Axt}},\ and\ \bibinfo {author}
  {\bibfnamefont {T.}~\bibnamefont {Kuhn}},\ }\bibfield  {title} {\bibinfo
  {title} {{Persistent Oscillations of the Order Parameter and Interaction
  Quench Phase Diagram for a Confined Bardeen-Cooper-Schrieffer Fermi Gas}},\
  }\href {https://doi.org/10.1103/PhysRevA.98.053605} {\bibfield  {journal}
  {\bibinfo  {journal} {Phys. Rev. A}\ }\textbf {\bibinfo {volume} {98}},\
  \bibinfo {pages} {053605} (\bibinfo {year} {2018})}\BibitemShut {NoStop}%
\bibitem [{\citenamefont {Seibold}\ and\ \citenamefont
  {Lorenzana}(2020)}]{Seibold2020PRB}%
  \BibitemOpen
  \bibfield  {author} {\bibinfo {author} {\bibfnamefont {G.}~\bibnamefont
  {Seibold}}\ and\ \bibinfo {author} {\bibfnamefont {J.}~\bibnamefont
  {Lorenzana}},\ }\bibfield  {title} {\bibinfo {title} {{Nonequilibrium
  Dynamics from BCS to the Bosonic Limit}},\ }\href
  {https://doi.org/10.1103/PhysRevB.102.144502} {\bibfield  {journal} {\bibinfo
   {journal} {Phys. Rev. B}\ }\textbf {\bibinfo {volume} {102}},\ \bibinfo
  {pages} {144502} (\bibinfo {year} {2020})}\BibitemShut {NoStop}%
\bibitem [{\citenamefont {Collado}\ \emph {et~al.}(2022)\citenamefont
  {Collado}, \citenamefont {Defenu},\ and\ \citenamefont
  {Lorenzana}}]{Collado2022arXiv}%
  \BibitemOpen
  \bibfield  {author} {\bibinfo {author} {\bibfnamefont {H.~P.}\ \bibnamefont
  {Collado}}, \bibinfo {author} {\bibfnamefont {N.}~\bibnamefont {Defenu}},\
  and\ \bibinfo {author} {\bibfnamefont {J.}~\bibnamefont {Lorenzana}},\
  }\bibfield  {title} {\bibinfo {title} {{Engineering Higgs Dynamics by
  Spectral Singularities}},\ }\href {https://arxiv.org/abs/2205.06826}
  {\bibfield  {journal} {\bibinfo  {journal} {arXiv:2205.06826}\ } (\bibinfo
  {year} {2022})}\BibitemShut {NoStop}%
\bibitem [{\citenamefont {Yi}\ and\ \citenamefont {Duan}(2006)}]{Yi2006PRA}%
  \BibitemOpen
  \bibfield  {author} {\bibinfo {author} {\bibfnamefont {W.}~\bibnamefont
  {Yi}}\ and\ \bibinfo {author} {\bibfnamefont {L.-M.}\ \bibnamefont {Duan}},\
  }\bibfield  {title} {\bibinfo {title} {{Dynamic Response of an Ultracold
  Fermi Gas near the Feshbach Resonance}},\ }\href
  {https://doi.org/10.1103/PhysRevA.73.013609} {\bibfield  {journal} {\bibinfo
  {journal} {Phys. Rev. A}\ }\textbf {\bibinfo {volume} {73}},\ \bibinfo
  {pages} {013609} (\bibinfo {year} {2006})}\BibitemShut {NoStop}%
\bibitem [{\citenamefont {Scott}\ \emph {et~al.}(2012)\citenamefont {Scott},
  \citenamefont {Dalfovo}, \citenamefont {Pitaevskii},\ and\ \citenamefont
  {Stringari}}]{Scott2012PRA}%
  \BibitemOpen
  \bibfield  {author} {\bibinfo {author} {\bibfnamefont {R.~G.}\ \bibnamefont
  {Scott}}, \bibinfo {author} {\bibfnamefont {F.}~\bibnamefont {Dalfovo}},
  \bibinfo {author} {\bibfnamefont {L.~P.}\ \bibnamefont {Pitaevskii}},\ and\
  \bibinfo {author} {\bibfnamefont {S.}~\bibnamefont {Stringari}},\ }\bibfield
  {title} {\bibinfo {title} {{Rapid Ramps Across the BEC-BCS crossover: A Route
  to Measuring the Superfluid Gap}},\ }\href
  {https://doi.org/10.1103/PhysRevA.86.053604} {\bibfield  {journal} {\bibinfo
  {journal} {Phys. Rev. A}\ }\textbf {\bibinfo {volume} {86}},\ \bibinfo
  {pages} {053604} (\bibinfo {year} {2012})}\BibitemShut {NoStop}%
\bibitem [{\citenamefont {Tokimoto}\ \emph {et~al.}(2019)\citenamefont
  {Tokimoto}, \citenamefont {Tsuchiya},\ and\ \citenamefont
  {Nikuni}}]{Tokimoto2019JPSJ}%
  \BibitemOpen
  \bibfield  {author} {\bibinfo {author} {\bibfnamefont {J.}~\bibnamefont
  {Tokimoto}}, \bibinfo {author} {\bibfnamefont {S.}~\bibnamefont {Tsuchiya}},\
  and\ \bibinfo {author} {\bibfnamefont {T.}~\bibnamefont {Nikuni}},\
  }\bibfield  {title} {\bibinfo {title} {{Excitation of Higgs Mode in
  Superfluid Fermi Gas in BCS–BEC Crossover}},\ }\href
  {https://doi.org/10.7566/JPSJ.88.023601} {\bibfield  {journal} {\bibinfo
  {journal} {J. Phys. Soc. Jpn.}\ }\textbf {\bibinfo {volume} {88}},\ \bibinfo
  {pages} {023601} (\bibinfo {year} {2019})}\BibitemShut {NoStop}%
\bibitem [{\citenamefont {Ojeda~Collado}\ \emph {et~al.}(2018)\citenamefont
  {Ojeda~Collado}, \citenamefont {Lorenzana}, \citenamefont {Usaj},\ and\
  \citenamefont {Balseiro}}]{Collado2018PRB}%
  \BibitemOpen
  \bibfield  {author} {\bibinfo {author} {\bibfnamefont {H.~P.}\ \bibnamefont
  {Ojeda~Collado}}, \bibinfo {author} {\bibfnamefont {J.}~\bibnamefont
  {Lorenzana}}, \bibinfo {author} {\bibfnamefont {G.}~\bibnamefont {Usaj}},\
  and\ \bibinfo {author} {\bibfnamefont {C.~A.}\ \bibnamefont {Balseiro}},\
  }\bibfield  {title} {\bibinfo {title} {{Population Inversion and Dynamical
  Phase Transitions in a Driven Superconductor}},\ }\href
  {https://doi.org/10.1103/PhysRevB.98.214519} {\bibfield  {journal} {\bibinfo
  {journal} {Phys. Rev. B}\ }\textbf {\bibinfo {volume} {98}},\ \bibinfo
  {pages} {214519} (\bibinfo {year} {2018})}\BibitemShut {NoStop}%
\bibitem [{\citenamefont {Ojeda~Collado}\ \emph {et~al.}(2020)\citenamefont
  {Ojeda~Collado}, \citenamefont {Usaj}, \citenamefont {Lorenzana},\ and\
  \citenamefont {Balseiro}}]{Collado2020PRB}%
  \BibitemOpen
  \bibfield  {author} {\bibinfo {author} {\bibfnamefont {H.~P.}\ \bibnamefont
  {Ojeda~Collado}}, \bibinfo {author} {\bibfnamefont {G.}~\bibnamefont {Usaj}},
  \bibinfo {author} {\bibfnamefont {J.}~\bibnamefont {Lorenzana}},\ and\
  \bibinfo {author} {\bibfnamefont {C.~A.}\ \bibnamefont {Balseiro}},\
  }\bibfield  {title} {\bibinfo {title} {{Nonlinear Dynamics of Driven
  Superconductors with Dissipation}},\ }\href
  {https://doi.org/10.1103/PhysRevB.101.054502} {\bibfield  {journal} {\bibinfo
   {journal} {Phys. Rev. B}\ }\textbf {\bibinfo {volume} {101}},\ \bibinfo
  {pages} {054502} (\bibinfo {year} {2020})}\BibitemShut {NoStop}%
\bibitem [{\citenamefont {Ojeda~Collado}\ \emph {et~al.}(2021)\citenamefont
  {Ojeda~Collado}, \citenamefont {Usaj}, \citenamefont {Balseiro},
  \citenamefont {Zanette},\ and\ \citenamefont {Lorenzana}}]{Collado2021PRR}%
  \BibitemOpen
  \bibfield  {author} {\bibinfo {author} {\bibfnamefont {H.~P.}\ \bibnamefont
  {Ojeda~Collado}}, \bibinfo {author} {\bibfnamefont {G.}~\bibnamefont {Usaj}},
  \bibinfo {author} {\bibfnamefont {C.~A.}\ \bibnamefont {Balseiro}}, \bibinfo
  {author} {\bibfnamefont {D.~H.}\ \bibnamefont {Zanette}},\ and\ \bibinfo
  {author} {\bibfnamefont {J.}~\bibnamefont {Lorenzana}},\ }\bibfield  {title}
  {\bibinfo {title} {{Emergent Parametric Resonances and Time-Crystal Phases in
  Driven Bardeen-Cooper-Schrieffer Systems}},\ }\href
  {https://doi.org/10.1103/PhysRevResearch.3.L042023} {\bibfield  {journal}
  {\bibinfo  {journal} {Phys. Rev. Res.}\ }\textbf {\bibinfo {volume} {3}},\
  \bibinfo {pages} {L042023} (\bibinfo {year} {2021})}\BibitemShut {NoStop}%
\bibitem [{\citenamefont {Bruun}(2002)}]{Bruun2002PRL}%
  \BibitemOpen
  \bibfield  {author} {\bibinfo {author} {\bibfnamefont {G.~M.}\ \bibnamefont
  {Bruun}},\ }\bibfield  {title} {\bibinfo {title} {{Low-Energy Monopole Modes
  of a Trapped Atomic Fermi Gas}},\ }\href
  {https://doi.org/10.1103/PhysRevLett.89.263002} {\bibfield  {journal}
  {\bibinfo  {journal} {Phys. Rev. Lett.}\ }\textbf {\bibinfo {volume} {89}},\
  \bibinfo {pages} {263002} (\bibinfo {year} {2002})}\BibitemShut {NoStop}%
\bibitem [{\citenamefont {Korolyuk}\ \emph {et~al.}(2011)\citenamefont
  {Korolyuk}, \citenamefont {Kinnunen},\ and\ \citenamefont
  {T\"orm\"a}}]{Korolyuk2011PRA}%
  \BibitemOpen
  \bibfield  {author} {\bibinfo {author} {\bibfnamefont {A.}~\bibnamefont
  {Korolyuk}}, \bibinfo {author} {\bibfnamefont {J.~J.}\ \bibnamefont
  {Kinnunen}},\ and\ \bibinfo {author} {\bibfnamefont {P.}~\bibnamefont
  {T\"orm\"a}},\ }\bibfield  {title} {\bibinfo {title} {{Density Response of a
  Trapped Fermi Gas: A Crossover from the Pair Vibration Mode to the Goldstone
  Mode}},\ }\href {https://doi.org/10.1103/PhysRevA.84.033623} {\bibfield
  {journal} {\bibinfo  {journal} {Phys. Rev. A}\ }\textbf {\bibinfo {volume}
  {84}},\ \bibinfo {pages} {033623} (\bibinfo {year} {2011})}\BibitemShut
  {NoStop}%
\bibitem [{\citenamefont {Podolsky}\ \emph {et~al.}(2011)\citenamefont
  {Podolsky}, \citenamefont {Auerbach},\ and\ \citenamefont
  {Arovas}}]{Podolsky2011PRB}%
  \BibitemOpen
  \bibfield  {author} {\bibinfo {author} {\bibfnamefont {D.}~\bibnamefont
  {Podolsky}}, \bibinfo {author} {\bibfnamefont {A.}~\bibnamefont {Auerbach}},\
  and\ \bibinfo {author} {\bibfnamefont {D.~P.}\ \bibnamefont {Arovas}},\
  }\bibfield  {title} {\bibinfo {title} {{Visibility of the Amplitude (Higgs)
  Mode in Condensed Matter}},\ }\href
  {https://doi.org/10.1103/PhysRevB.84.174522} {\bibfield  {journal} {\bibinfo
  {journal} {Phys. Rev. B}\ }\textbf {\bibinfo {volume} {84}},\ \bibinfo
  {pages} {174522} (\bibinfo {year} {2011})}\BibitemShut {NoStop}%
\bibitem [{\citenamefont {Liu}\ \emph {et~al.}(2016)\citenamefont {Liu},
  \citenamefont {Zhai},\ and\ \citenamefont {Zhang}}]{Liu2016PRA}%
  \BibitemOpen
  \bibfield  {author} {\bibinfo {author} {\bibfnamefont {B.}~\bibnamefont
  {Liu}}, \bibinfo {author} {\bibfnamefont {H.}~\bibnamefont {Zhai}},\ and\
  \bibinfo {author} {\bibfnamefont {S.}~\bibnamefont {Zhang}},\ }\bibfield
  {title} {\bibinfo {title} {{Evolution of the Higgs Mode in a Fermion
  Superfluid with Tunable Interactions}},\ }\href
  {https://doi.org/10.1103/PhysRevA.93.033641} {\bibfield  {journal} {\bibinfo
  {journal} {Phys. Rev. A}\ }\textbf {\bibinfo {volume} {93}},\ \bibinfo
  {pages} {033641} (\bibinfo {year} {2016})}\BibitemShut {NoStop}%
\bibitem [{\citenamefont {Han}\ \emph {et~al.}(2016)\citenamefont {Han},
  \citenamefont {Liu},\ and\ \citenamefont {Hu}}]{Han2016PRA}%
  \BibitemOpen
  \bibfield  {author} {\bibinfo {author} {\bibfnamefont {X.}~\bibnamefont
  {Han}}, \bibinfo {author} {\bibfnamefont {B.}~\bibnamefont {Liu}},\ and\
  \bibinfo {author} {\bibfnamefont {J.}~\bibnamefont {Hu}},\ }\bibfield
  {title} {\bibinfo {title} {{Observability of Higgs Mode in a System without
  Lorentz Invariance}},\ }\href {https://doi.org/10.1103/PhysRevA.94.033608}
  {\bibfield  {journal} {\bibinfo  {journal} {Phys. Rev. A}\ }\textbf {\bibinfo
  {volume} {94}},\ \bibinfo {pages} {033608} (\bibinfo {year}
  {2016})}\BibitemShut {NoStop}%
\bibitem [{\citenamefont {Varma}(2002)}]{Varma2002JLTP}%
  \BibitemOpen
  \bibfield  {author} {\bibinfo {author} {\bibfnamefont {C.~M.}\ \bibnamefont
  {Varma}},\ }\bibfield  {title} {\bibinfo {title} {{Higgs Boson in
  Superconductors}},\ }\href {https://doi.org/10.1023/A:1013890507658}
  {\bibfield  {journal} {\bibinfo  {journal} {J. Low Temp. Phys.}\ }\textbf
  {\bibinfo {volume} {126}},\ \bibinfo {pages} {901} (\bibinfo {year}
  {2002})}\BibitemShut {NoStop}%
\bibitem [{\citenamefont {Akbari}\ \emph {et~al.}(2013)\citenamefont {Akbari},
  \citenamefont {Schnyder}, \citenamefont {Manske},\ and\ \citenamefont
  {Eremin}}]{Akbari2013EPL}%
  \BibitemOpen
  \bibfield  {author} {\bibinfo {author} {\bibfnamefont {A.}~\bibnamefont
  {Akbari}}, \bibinfo {author} {\bibfnamefont {A.~P.}\ \bibnamefont
  {Schnyder}}, \bibinfo {author} {\bibfnamefont {D.}~\bibnamefont {Manske}},\
  and\ \bibinfo {author} {\bibfnamefont {I.}~\bibnamefont {Eremin}},\
  }\bibfield  {title} {\bibinfo {title} {{Theory of Nonequilibrium Dynamics of
  Multiband Superconductors}},\ }\href
  {https://doi.org/10.1209/0295-5075/101/17002} {\bibfield  {journal} {\bibinfo
   {journal} {Europhysics Letters}\ }\textbf {\bibinfo {volume} {101}},\
  \bibinfo {pages} {17002} (\bibinfo {year} {2013})}\BibitemShut {NoStop}%
\bibitem [{\citenamefont {Maiti}\ and\ \citenamefont
  {Hirschfeld}(2015)}]{Maiti2015PRB}%
  \BibitemOpen
  \bibfield  {author} {\bibinfo {author} {\bibfnamefont {S.}~\bibnamefont
  {Maiti}}\ and\ \bibinfo {author} {\bibfnamefont {P.~J.}\ \bibnamefont
  {Hirschfeld}},\ }\bibfield  {title} {\bibinfo {title} {{Collective Modes in
  Superconductors with Competing $s$- and $d$-Wave Interactions}},\ }\href
  {https://doi.org/10.1103/PhysRevB.92.094506} {\bibfield  {journal} {\bibinfo
  {journal} {Phys. Rev. B}\ }\textbf {\bibinfo {volume} {92}},\ \bibinfo
  {pages} {094506} (\bibinfo {year} {2015})}\BibitemShut {NoStop}%
\bibitem [{\citenamefont {M\"uller}\ \emph {et~al.}(2018)\citenamefont
  {M\"uller}, \citenamefont {Shen}, \citenamefont {Dzero},\ and\ \citenamefont
  {Eremin}}]{Muller2018PRB}%
  \BibitemOpen
  \bibfield  {author} {\bibinfo {author} {\bibfnamefont {M.~A.}\ \bibnamefont
  {M\"uller}}, \bibinfo {author} {\bibfnamefont {P.}~\bibnamefont {Shen}},
  \bibinfo {author} {\bibfnamefont {M.}~\bibnamefont {Dzero}},\ and\ \bibinfo
  {author} {\bibfnamefont {I.}~\bibnamefont {Eremin}},\ }\bibfield  {title}
  {\bibinfo {title} {{Short-Time Dynamics in $s+is$-Wave Superconductor with
  Incipient Bands}},\ }\href {https://doi.org/10.1103/PhysRevB.98.024522}
  {\bibfield  {journal} {\bibinfo  {journal} {Phys. Rev. B}\ }\textbf {\bibinfo
  {volume} {98}},\ \bibinfo {pages} {024522} (\bibinfo {year}
  {2018})}\BibitemShut {NoStop}%
\bibitem [{\citenamefont {Fausti}\ \emph {et~al.}(2011)\citenamefont {Fausti},
  \citenamefont {Tobey}, \citenamefont {Dean}, \citenamefont {Kaiser},
  \citenamefont {Dienst}, \citenamefont {Hoffmann}, \citenamefont {Pyon},
  \citenamefont {Takayama}, \citenamefont {Takagi},\ and\ \citenamefont
  {Cavalleri}}]{Fausti2011science}%
  \BibitemOpen
  \bibfield  {author} {\bibinfo {author} {\bibfnamefont {D.}~\bibnamefont
  {Fausti}}, \bibinfo {author} {\bibfnamefont {R.}~\bibnamefont {Tobey}},
  \bibinfo {author} {\bibfnamefont {N.}~\bibnamefont {Dean}}, \bibinfo {author}
  {\bibfnamefont {S.}~\bibnamefont {Kaiser}}, \bibinfo {author} {\bibfnamefont
  {A.}~\bibnamefont {Dienst}}, \bibinfo {author} {\bibfnamefont {M.~C.}\
  \bibnamefont {Hoffmann}}, \bibinfo {author} {\bibfnamefont {S.}~\bibnamefont
  {Pyon}}, \bibinfo {author} {\bibfnamefont {T.}~\bibnamefont {Takayama}},
  \bibinfo {author} {\bibfnamefont {H.}~\bibnamefont {Takagi}},\ and\ \bibinfo
  {author} {\bibfnamefont {A.}~\bibnamefont {Cavalleri}},\ }\bibfield  {title}
  {\bibinfo {title} {{Light-Induced Superconductivity in a Stripe-Ordered
  Cuprate}},\ }\href {https://www.science.org/doi/abs/10.1126/science.1197294}
  {\bibfield  {journal} {\bibinfo  {journal} {science}\ }\textbf {\bibinfo
  {volume} {331}},\ \bibinfo {pages} {189} (\bibinfo {year}
  {2011})}\BibitemShut {NoStop}%
\bibitem [{\citenamefont {Kaiser}\ \emph {et~al.}(2014)\citenamefont {Kaiser},
  \citenamefont {Hunt}, \citenamefont {Nicoletti}, \citenamefont {Hu},
  \citenamefont {Gierz}, \citenamefont {Liu}, \citenamefont {Le~Tacon},
  \citenamefont {Loew}, \citenamefont {Haug}, \citenamefont {Keimer},\ and\
  \citenamefont {Cavalleri}}]{Kaiser2014PRB}%
  \BibitemOpen
  \bibfield  {author} {\bibinfo {author} {\bibfnamefont {S.}~\bibnamefont
  {Kaiser}}, \bibinfo {author} {\bibfnamefont {C.~R.}\ \bibnamefont {Hunt}},
  \bibinfo {author} {\bibfnamefont {D.}~\bibnamefont {Nicoletti}}, \bibinfo
  {author} {\bibfnamefont {W.}~\bibnamefont {Hu}}, \bibinfo {author}
  {\bibfnamefont {I.}~\bibnamefont {Gierz}}, \bibinfo {author} {\bibfnamefont
  {H.~Y.}\ \bibnamefont {Liu}}, \bibinfo {author} {\bibfnamefont
  {M.}~\bibnamefont {Le~Tacon}}, \bibinfo {author} {\bibfnamefont
  {T.}~\bibnamefont {Loew}}, \bibinfo {author} {\bibfnamefont {D.}~\bibnamefont
  {Haug}}, \bibinfo {author} {\bibfnamefont {B.}~\bibnamefont {Keimer}},\ and\
  \bibinfo {author} {\bibfnamefont {A.}~\bibnamefont {Cavalleri}},\ }\bibfield
  {title} {\bibinfo {title} {{Optically Induced Coherent Transport Far above
  \text{${T}_{c}$} in Underdoped
  \text{${\mathrm{YBa}}_{2}{\mathrm{Cu}}_{3}{\mathrm{O}}_{6+\ensuremath{\delta}}$}
  }},\ }\href {https://doi.org/10.1103/PhysRevB.89.184516} {\bibfield
  {journal} {\bibinfo  {journal} {Phys. Rev. B}\ }\textbf {\bibinfo {volume}
  {89}},\ \bibinfo {pages} {184516} (\bibinfo {year} {2014})}\BibitemShut
  {NoStop}%
\bibitem [{\citenamefont {Hu}\ \emph {et~al.}(2014)\citenamefont {Hu},
  \citenamefont {Kaiser}, \citenamefont {Nicoletti}, \citenamefont {Hunt},
  \citenamefont {Gierz}, \citenamefont {Hoffmann}, \citenamefont {Le~Tacon},
  \citenamefont {Loew}, \citenamefont {Keimer},\ and\ \citenamefont
  {Cavalleri}}]{Hu2014naturemat}%
  \BibitemOpen
  \bibfield  {author} {\bibinfo {author} {\bibfnamefont {W.}~\bibnamefont
  {Hu}}, \bibinfo {author} {\bibfnamefont {S.}~\bibnamefont {Kaiser}}, \bibinfo
  {author} {\bibfnamefont {D.}~\bibnamefont {Nicoletti}}, \bibinfo {author}
  {\bibfnamefont {C.~R.}\ \bibnamefont {Hunt}}, \bibinfo {author}
  {\bibfnamefont {I.}~\bibnamefont {Gierz}}, \bibinfo {author} {\bibfnamefont
  {M.~C.}\ \bibnamefont {Hoffmann}}, \bibinfo {author} {\bibfnamefont
  {M.}~\bibnamefont {Le~Tacon}}, \bibinfo {author} {\bibfnamefont
  {T.}~\bibnamefont {Loew}}, \bibinfo {author} {\bibfnamefont {B.}~\bibnamefont
  {Keimer}},\ and\ \bibinfo {author} {\bibfnamefont {A.}~\bibnamefont
  {Cavalleri}},\ }\bibfield  {title} {\bibinfo {title} {{Optically Enhanced
  Coherent Transport in
  \text{${\mathrm{YBa}}_{2}{\mathrm{Cu}}_{3}{\mathrm{O}}_{6.5}$} by Ultrafast
  Redistribution of Interlayer Coupling}},\ }\href
  {https://doi.org/10.1038/nmat3963} {\bibfield  {journal} {\bibinfo  {journal}
  {Nature materials}\ }\textbf {\bibinfo {volume} {13}},\ \bibinfo {pages}
  {705} (\bibinfo {year} {2014})}\BibitemShut {NoStop}%
\bibitem [{\citenamefont {Mitrano}\ \emph {et~al.}(2016)\citenamefont
  {Mitrano}, \citenamefont {Cantaluppi}, \citenamefont {Nicoletti},
  \citenamefont {Kaiser}, \citenamefont {Perucchi}, \citenamefont {Lupi},
  \citenamefont {Di~Pietro}, \citenamefont {Pontiroli}, \citenamefont
  {Ricc{\`o}}, \citenamefont {Clark}, \citenamefont {Jaksch},\ and\
  \citenamefont {Cavalleri}}]{Mitrano2016nature}%
  \BibitemOpen
  \bibfield  {author} {\bibinfo {author} {\bibfnamefont {M.}~\bibnamefont
  {Mitrano}}, \bibinfo {author} {\bibfnamefont {A.}~\bibnamefont {Cantaluppi}},
  \bibinfo {author} {\bibfnamefont {D.}~\bibnamefont {Nicoletti}}, \bibinfo
  {author} {\bibfnamefont {S.}~\bibnamefont {Kaiser}}, \bibinfo {author}
  {\bibfnamefont {A.}~\bibnamefont {Perucchi}}, \bibinfo {author}
  {\bibfnamefont {S.}~\bibnamefont {Lupi}}, \bibinfo {author} {\bibfnamefont
  {P.}~\bibnamefont {Di~Pietro}}, \bibinfo {author} {\bibfnamefont
  {D.}~\bibnamefont {Pontiroli}}, \bibinfo {author} {\bibfnamefont
  {M.}~\bibnamefont {Ricc{\`o}}}, \bibinfo {author} {\bibfnamefont {S.~R.}\
  \bibnamefont {Clark}}, \bibinfo {author} {\bibfnamefont {D.}~\bibnamefont
  {Jaksch}},\ and\ \bibinfo {author} {\bibfnamefont {A.}~\bibnamefont
  {Cavalleri}},\ }\bibfield  {title} {\bibinfo {title} {{Possible Light-Induced
  Superconductivity in \text{$\mathrm{K}_3\mathrm{C}_{60}$} at High
  Temperature}},\ }\href {https://doi.org/10.1038/nature16522} {\bibfield
  {journal} {\bibinfo  {journal} {Nature}\ }\textbf {\bibinfo {volume} {530}},\
  \bibinfo {pages} {461} (\bibinfo {year} {2016})}\BibitemShut {NoStop}%
\bibitem [{\citenamefont {Bruun}(2014)}]{Bruun2014PRA}%
  \BibitemOpen
  \bibfield  {author} {\bibinfo {author} {\bibfnamefont {G.~M.}\ \bibnamefont
  {Bruun}},\ }\bibfield  {title} {\bibinfo {title} {{Long-lived Higgs Mode in a
  Two-Dimensional Confined Fermi System}},\ }\href
  {https://doi.org/10.1103/PhysRevA.90.023621} {\bibfield  {journal} {\bibinfo
  {journal} {Phys. Rev. A}\ }\textbf {\bibinfo {volume} {90}},\ \bibinfo
  {pages} {023621} (\bibinfo {year} {2014})}\BibitemShut {NoStop}%
\bibitem [{\citenamefont {de~Gennes}(1966)}]{deGennes1966book}%
  \BibitemOpen
  \bibfield  {author} {\bibinfo {author} {\bibfnamefont {P.~G.}\ \bibnamefont
  {de~Gennes}},\ }\href {https://doi.org/10.1201/9780429497032} {\emph
  {\bibinfo {title} {{Superconductivity of Metals and Alloys}}}}\ (\bibinfo
  {publisher} {CRC Press},\ \bibinfo {address} {Boca Raton, FL},\ \bibinfo
  {year} {1966})\BibitemShut {NoStop}%
\bibitem [{\citenamefont {K{\"u}mmel}(1969)}]{Kummel1969ZPhysA}%
  \BibitemOpen
  \bibfield  {author} {\bibinfo {author} {\bibfnamefont {R.}~\bibnamefont
  {K{\"u}mmel}},\ }\bibfield  {title} {\bibinfo {title} {{Dynamics of Current
  Flow Through the Phase-Boundary between a Normal and a Superconducting
  Region}},\ }\href {https://doi.org/10.1007/BF01392426} {\bibfield  {journal}
  {\bibinfo  {journal} {Z. Phys. A}\ }\textbf {\bibinfo {volume} {218}},\
  \bibinfo {pages} {472} (\bibinfo {year} {1969})}\BibitemShut {NoStop}%
\bibitem [{\citenamefont {Ketterson}\ and\ \citenamefont
  {Song}(1999)}]{Ketterson1999book}%
  \BibitemOpen
  \bibfield  {author} {\bibinfo {author} {\bibfnamefont {J.~B.}\ \bibnamefont
  {Ketterson}}\ and\ \bibinfo {author} {\bibfnamefont {S.~N.}\ \bibnamefont
  {Song}},\ }\href {https://doi.org/10.1017/CBO9781139171090} {\emph {\bibinfo
  {title} {Superconductivity}}}\ (\bibinfo  {publisher} {Cambridge University
  Press},\ \bibinfo {address} {Cambridge},\ \bibinfo {year} {1999})\BibitemShut
  {NoStop}%
\bibitem [{\citenamefont {Leggett}(1980)}]{Leggett1980proc}%
  \BibitemOpen
  \bibfield  {author} {\bibinfo {author} {\bibfnamefont {A.~J.}\ \bibnamefont
  {Leggett}},\ }\bibfield  {title} {\bibinfo {title} {Diatomic {M}olecules and
  {C}ooper {P}airs},\ }in\ \href@noop {} {\emph {\bibinfo {booktitle} {Modern
  Trends in the Theory of Condensed Matter}}},\ \bibinfo {editor} {edited by\
  \bibinfo {editor} {\bibfnamefont {A.}~\bibnamefont {P\k{e}kalski}}\ and\
  \bibinfo {editor} {\bibfnamefont {J.~A.}\ \bibnamefont {Przystawa}}}\
  (\bibinfo  {publisher} {Springer Berlin Heidelberg},\ \bibinfo {address}
  {Berlin, Heidelberg},\ \bibinfo {year} {1980})\ pp.\ \bibinfo {pages}
  {13--27}\BibitemShut {NoStop}%
\bibitem [{\citenamefont {Randeria}(1995)}]{Randeria1995book}%
  \BibitemOpen
  \bibfield  {author} {\bibinfo {author} {\bibfnamefont {M.}~\bibnamefont
  {Randeria}},\ }\bibinfo {title} {{Crossover from BCS Theory to
  Bose–Einstein Condensation}},\ in\ \href
  {https://doi.org/10.1017/CBO9780511524240.017} {\emph {\bibinfo {booktitle}
  {Bose-Einstein Condensation}}},\ \bibinfo {editor} {edited by\ \bibinfo
  {editor} {\bibfnamefont {A.}~\bibnamefont {Griffin}}, \bibinfo {editor}
  {\bibfnamefont {D.~W.}\ \bibnamefont {Snoke}},\ and\ \bibinfo {editor}
  {\bibfnamefont {S.}~\bibnamefont {Stringari}}}\ (\bibinfo  {publisher}
  {Cambridge University Press},\ \bibinfo {address} {Cambridge},\ \bibinfo
  {year} {1995})\ p.\ \bibinfo {pages} {355–392}\BibitemShut {NoStop}%
\bibitem [{\citenamefont {Giorgini}\ \emph {et~al.}(2008)\citenamefont
  {Giorgini}, \citenamefont {Pitaevskii},\ and\ \citenamefont
  {Stringari}}]{Giorgini2008RMP}%
  \BibitemOpen
  \bibfield  {author} {\bibinfo {author} {\bibfnamefont {S.}~\bibnamefont
  {Giorgini}}, \bibinfo {author} {\bibfnamefont {L.~P.}\ \bibnamefont
  {Pitaevskii}},\ and\ \bibinfo {author} {\bibfnamefont {S.}~\bibnamefont
  {Stringari}},\ }\bibfield  {title} {\bibinfo {title} {{Theory of Ultracold
  Atomic Fermi Gases}},\ }\href {https://doi.org/10.1103/RevModPhys.80.1215}
  {\bibfield  {journal} {\bibinfo  {journal} {Rev. Mod. Phys.}\ }\textbf
  {\bibinfo {volume} {80}},\ \bibinfo {pages} {1215} (\bibinfo {year}
  {2008})}\BibitemShut {NoStop}%
\bibitem [{\citenamefont {Greiner}\ \emph {et~al.}(2005)\citenamefont
  {Greiner}, \citenamefont {Regal},\ and\ \citenamefont
  {Jin}}]{Greiner2005PRL}%
  \BibitemOpen
  \bibfield  {author} {\bibinfo {author} {\bibfnamefont {M.}~\bibnamefont
  {Greiner}}, \bibinfo {author} {\bibfnamefont {C.~A.}\ \bibnamefont {Regal}},\
  and\ \bibinfo {author} {\bibfnamefont {D.~S.}\ \bibnamefont {Jin}},\
  }\bibfield  {title} {\bibinfo {title} {{Probing the Excitation Spectrum of a
  Fermi Gas in the BCS-BEC Crossover Regime}},\ }\href
  {https://doi.org/10.1103/PhysRevLett.94.070403} {\bibfield  {journal}
  {\bibinfo  {journal} {Phys. Rev. Lett.}\ }\textbf {\bibinfo {volume} {94}},\
  \bibinfo {pages} {070403} (\bibinfo {year} {2005})}\BibitemShut {NoStop}%
\bibitem [{\citenamefont {Thompson}\ \emph {et~al.}(2005)\citenamefont
  {Thompson}, \citenamefont {Hodby},\ and\ \citenamefont
  {Wieman}}]{Thompson2005PRL}%
  \BibitemOpen
  \bibfield  {author} {\bibinfo {author} {\bibfnamefont {S.~T.}\ \bibnamefont
  {Thompson}}, \bibinfo {author} {\bibfnamefont {E.}~\bibnamefont {Hodby}},\
  and\ \bibinfo {author} {\bibfnamefont {C.~E.}\ \bibnamefont {Wieman}},\
  }\bibfield  {title} {\bibinfo {title} {{Ultracold Molecule Production via a
  Resonant Oscillating Magnetic Field}},\ }\href
  {https://doi.org/10.1103/PhysRevLett.95.190404} {\bibfield  {journal}
  {\bibinfo  {journal} {Phys. Rev. Lett.}\ }\textbf {\bibinfo {volume} {95}},\
  \bibinfo {pages} {190404} (\bibinfo {year} {2005})}\BibitemShut {NoStop}%
\bibitem [{\citenamefont {Clark}\ \emph {et~al.}(2017)\citenamefont {Clark},
  \citenamefont {Gaj}, \citenamefont {Feng},\ and\ \citenamefont
  {Chin}}]{Clark2017nature}%
  \BibitemOpen
  \bibfield  {author} {\bibinfo {author} {\bibfnamefont {L.~W.}\ \bibnamefont
  {Clark}}, \bibinfo {author} {\bibfnamefont {A.}~\bibnamefont {Gaj}}, \bibinfo
  {author} {\bibfnamefont {L.}~\bibnamefont {Feng}},\ and\ \bibinfo {author}
  {\bibfnamefont {C.}~\bibnamefont {Chin}},\ }\bibfield  {title} {\bibinfo
  {title} {{Collective Emission of Matter-Wave Jets from Driven Bose--Einstein
  Condensates}},\ }\href {https://doi.org/10.1038/nature24272} {\bibfield
  {journal} {\bibinfo  {journal} {Nature (London)}\ }\textbf {\bibinfo {volume}
  {551}},\ \bibinfo {pages} {356} (\bibinfo {year} {2017})}\BibitemShut
  {NoStop}%
\bibitem [{\citenamefont {Chin}\ \emph {et~al.}(2010)\citenamefont {Chin},
  \citenamefont {Grimm}, \citenamefont {Julienne},\ and\ \citenamefont
  {Tiesinga}}]{Chin2010RMP}%
  \BibitemOpen
  \bibfield  {author} {\bibinfo {author} {\bibfnamefont {C.}~\bibnamefont
  {Chin}}, \bibinfo {author} {\bibfnamefont {R.}~\bibnamefont {Grimm}},
  \bibinfo {author} {\bibfnamefont {P.}~\bibnamefont {Julienne}},\ and\
  \bibinfo {author} {\bibfnamefont {E.}~\bibnamefont {Tiesinga}},\ }\bibfield
  {title} {\bibinfo {title} {{Feshbach Resonances in Ultracold Gases}},\ }\href
  {https://doi.org/10.1103/RevModPhys.82.1225} {\bibfield  {journal} {\bibinfo
  {journal} {Rev. Mod. Phys.}\ }\textbf {\bibinfo {volume} {82}},\ \bibinfo
  {pages} {1225} (\bibinfo {year} {2010})}\BibitemShut {NoStop}%
\bibitem [{\citenamefont {Yariv}(1988)}]{Yariv1988book}%
  \BibitemOpen
  \bibfield  {author} {\bibinfo {author} {\bibfnamefont {A.}~\bibnamefont
  {Yariv}},\ }\href
  {https://www.wiley.com/en-gb/Quantum+Electronics%2C+3rd+Edition-p-9780471609971#evaluation-copy-section}
  {\emph {\bibinfo {title} {{Quantum Electronics}}}},\ \bibinfo {edition}
  {3rd}\ ed.\ (\bibinfo  {publisher} {Wiley},\ \bibinfo {address} {New York},\
  \bibinfo {year} {1988})\BibitemShut {NoStop}%
\bibitem [{\citenamefont {Porsev}\ and\ \citenamefont
  {Derevianko}(2004)}]{Porsev2004PRA}%
  \BibitemOpen
  \bibfield  {author} {\bibinfo {author} {\bibfnamefont {S.~G.}\ \bibnamefont
  {Porsev}}\ and\ \bibinfo {author} {\bibfnamefont {A.}~\bibnamefont
  {Derevianko}},\ }\bibfield  {title} {\bibinfo {title} {{Hyperfine Quenching
  of the Metastable \text{$^{3}P_{0,2}$} states in Divalent Atoms}},\ }\href
  {https://doi.org/10.1103/PhysRevA.69.042506} {\bibfield  {journal} {\bibinfo
  {journal} {Phys. Rev. A}\ }\textbf {\bibinfo {volume} {69}},\ \bibinfo
  {pages} {042506} (\bibinfo {year} {2004})}\BibitemShut {NoStop}%
\bibitem [{\citenamefont {Fukuhara}\ \emph {et~al.}(2007)\citenamefont
  {Fukuhara}, \citenamefont {Takasu}, \citenamefont {Kumakura},\ and\
  \citenamefont {Takahashi}}]{Fukuhara2007PRL}%
  \BibitemOpen
  \bibfield  {author} {\bibinfo {author} {\bibfnamefont {T.}~\bibnamefont
  {Fukuhara}}, \bibinfo {author} {\bibfnamefont {Y.}~\bibnamefont {Takasu}},
  \bibinfo {author} {\bibfnamefont {M.}~\bibnamefont {Kumakura}},\ and\
  \bibinfo {author} {\bibfnamefont {Y.}~\bibnamefont {Takahashi}},\ }\bibfield
  {title} {\bibinfo {title} {{Degenerate Fermi Gases of Ytterbium}},\ }\href
  {https://doi.org/10.1103/PhysRevLett.98.030401} {\bibfield  {journal}
  {\bibinfo  {journal} {Phys. Rev. Lett.}\ }\textbf {\bibinfo {volume} {98}},\
  \bibinfo {pages} {030401} (\bibinfo {year} {2007})}\BibitemShut {NoStop}%
\bibitem [{\citenamefont {Gaunt}\ \emph {et~al.}(2013)\citenamefont {Gaunt},
  \citenamefont {Schmidutz}, \citenamefont {Gotlibovych}, \citenamefont
  {Smith},\ and\ \citenamefont {Hadzibabic}}]{Gaunt2013PRL}%
  \BibitemOpen
  \bibfield  {author} {\bibinfo {author} {\bibfnamefont {A.~L.}\ \bibnamefont
  {Gaunt}}, \bibinfo {author} {\bibfnamefont {T.~F.}\ \bibnamefont
  {Schmidutz}}, \bibinfo {author} {\bibfnamefont {I.}~\bibnamefont
  {Gotlibovych}}, \bibinfo {author} {\bibfnamefont {R.~P.}\ \bibnamefont
  {Smith}},\ and\ \bibinfo {author} {\bibfnamefont {Z.}~\bibnamefont
  {Hadzibabic}},\ }\bibfield  {title} {\bibinfo {title} {{Bose-Einstein
  Condensation of Atoms in a Uniform Potential}},\ }\href
  {https://doi.org/10.1103/PhysRevLett.110.200406} {\bibfield  {journal}
  {\bibinfo  {journal} {Phys. Rev. Lett.}\ }\textbf {\bibinfo {volume} {110}},\
  \bibinfo {pages} {200406} (\bibinfo {year} {2013})}\BibitemShut {NoStop}%
\end{thebibliography}%

\end{document}